\documentclass[a4paper,12pt]{article}
\usepackage[margin=2.5cm]{geometry}
\usepackage{amsmath}
\usepackage{mathtools}
\usepackage{physics}
\usepackage[utf8]{inputenc}
\usepackage{multirow}
\usepackage{amssymb}
\usepackage{authblk}
\usepackage{graphicx,amsfonts}
\usepackage[sort&compress,numbers]{natbib}
\usepackage{float}
\usepackage[svgnames]{xcolor}
\usepackage{chngcntr}
\usepackage{caption}
\usepackage{subcaption}
\usepackage{comment}
\usepackage{ulem}
\counterwithin{figure}{section}
\usepackage[colorlinks,citecolor=blue,urlcolor=blue,linkcolor=blue]{hyperref}

\newcommand\be{\begin{equation}}
	\newcommand\ee{\end{equation}}
\newcommand\bea{\begin{eqnarray}}
	\newcommand\eea{\end{eqnarray}}

\title{Dyonic Taub-NUT-AdS Black Branes:\\
	Thermodynamics and Phase Diagrams}
\author[a]{Amr AlBarqawy\thanks{amr.albarqawy@sci.asu.edu.eg}}
\author[a,c]{Adel Awad \thanks{a.awad@sci.asu.edu.eg}}
\author[a]{Esraa Elkhateeb\thanks{dr.esraali@sci.asu.edu.eg}}
\author[b]{Mohamed Tharwat\thanks{oweg@aucegypt.edu}}

\affil[a]{\footnotesize \it Department of Physics,
	Faculty of Science, Ain Shams University, Cairo 11566, Egypt}
\affil[b]{\footnotesize \it Department of Physics, School of Sciences and Engineering, American University in Cairo, P.O. Box 74, AUC Avenue New Cairo, Cairo, Egypt}
\affil[c]{\footnotesize \it Centre for Theoretical Physics, the British University in Egypt, El Sherouk City 11837, Egypt}
\date{}

\begin{document}
	\maketitle
\begin{abstract}
Motivated by the recent developments in the thermodynamics of Taub-NUT spaces and the absence of Misner strings in Taub-NUT solutions with flat horizons, we study the phase structure of the dyonic Taub-NUT solutions with Lorentzian signature. We follow the treatment proposed in arXiv:2206.09124 and arXiv:2304.06705 to introduce the NUT parameter as a conserved charge to the first law. Although the calculated quantities satisfy the first law, we have found a larger class of charges that satisfy the first law and depend on some arbitrary parameter $\alpha$. We choose to describe phase diagrams as NUT parameter-Temperature graphs to show borders of big and small black brane phases. We study the phase structure of these spaces in a mixed ensemble (i.e., fixed electric potential, NUT charge, and magnetic charge), which we classify into different cases depending on the value of $\alpha$ and the other quantities. Most of these cases not only have first-order phase transitions, but also, end at critical points. Some of these cases include up to four critical points, depending on the value of $\alpha$ and the other quantities.

\end{abstract}
	\section{Introduction}
One of the most attractive features of anti-de Sitter space solutions is their positive specific heat \cite{Hawking:1982dh}. This feature reflects thermal stability of the system in contrast with the known negative specific heat of asymptotically flat spacetimes. In their seminal analysis of Schwarzschild-AdS black holes, Hawking and Page \cite{Hawking:1982dh} were able to show the possibility of phase transitions between the black hole and the pure thermal AdS phases. Upon discovering the AdS/CFT correspondence \cite{Maldacena:1997re,Witten:1998qj}, this transition was interpreted as the confinement and deconfinement phase transition of the dual gauge field theory. \\
\hfill\\
In short, the AdS/CFT correspondence is a duality that relates an ($n+1$)–dimensional theory of gravity on anti–de Sitter (AdS) spacetime to a gauge/conformal field theory (CFT) in the boundary that has $n$ dimensions. In this duality, there is a dictionary linking every quantity in the bulk (AdS-gravity) to another in the boundary (quantum field theory). For example, a mass in Schwarzschild-AdS is linked to a finite temperature on the boundary field theory. \\
\hfill\\
In 1999 Hawking-Page analysis was extended to the AdS charged black hole in \cite{Chamblin:1999tk,Chamblin:1999hg}, where the authors found first-order phase transitions, which end at a critical point. This transition is analogous to the well-known liquid-gas phase transition, or Van der Waals fluid. Later, many authors studied the phase structure of various charged AdS black holes, either by considering a fixed cosmological constant (Check for example, Refs. \cite{Dehyadegari:2016nkd,Dehyadegari:2017hvd,Hartnoll:2008kx,Hartnoll:2008vx}), or allowing it to vary, which is known as extended phase space thermodynamics (Check for example, Refs. \cite{Kubiznak:2012wp,Dutta:2013dca,Hennigar:2017umz,Tharwat:2023vku}). \\
\hfill\\
Working in AdS spaces enlarges the spectrum of various types of black holes, since they come with three horizon topologies; spherical, flat, or hyperbolic. Therefore, it is interesting to study the thermodynamics of AdS black holes with a flat horizon, which was investigated in \cite{Vanzo:1997gw, Birmingham:1998nr,Brill:1997mf}. Furthermore, their extended thermodynamics was studied in \cite{Dutta:2013dca} as well. However, these investigations showed no sign of first-order phase transitions or critical behavior since the black hole horizon temperature varies monotonically as a function of the horizon radius! However, several authors were able to show the existence of phase transition and critical behavior for these solutions by adding more exotic ingredients to the setup \cite{Hartnoll:2008kx,Hartnoll:2008vx,Surya:2001vj,bueno_holographic_2018,Dutta:2013dca,plantz_black_nodate,Hennigar:2017umz,Horowitz:2009ij}. In \cite{Hartnoll:2008vx, Hartnoll:2008kx,Horowitz:2009ij}, for example, the authors introduced a complex scalar field that gave rise to a superconductor phase diagram with a second-order phase transition between a black brane with a scalar hair at low temperature and a black brane with no hair at high temperature. Alternatively, in \cite{Hennigar:2017umz}, the authors achieved phase transition and critical behavior by introducing higher curvature terms with certain topological features to the gravitational action.\\
\hfill\\
One of the attractive AdS solutions studied in the above contexts is the AdS-Taub-Bolt/NUT black holes in four and higher dimensions \cite{Chamblin:1998pz,Emparan:1999pm,Page:1985bq,Awad:2000gg}. This class of solutions is a generalization to the Schwarzschild-AdS space with one additional parameter, the NUT parameter "n". An important feature of these solutions is that they possess string-like singularities known as Misner strings \cite{Misner:1963fr}. These strings are analogous to the Dirac string in electromagnetism and can be removed by imposing certain periodicity conditions, namely; the inverse temperature set to be $\beta =8\pi n$.\\
 \hfill\\
The Taub-NUT solution is often considered to be a gravitaional dyon \cite{Dowker:1974znr}, where the dual to the Komar mass is non-vanishing. This dual quantity defines the NUT charge $N$, interpreted as the gravitational analogue of magnetic charge. For Taub-NUT-AdS solution, the conserved charge $N$ can be calculated by taking the dual of the generalized Komar integral \cite{Kastor:2009wy, Bordo:2019tyh}
\begin{equation*}\label{con-N-int}
    N =  \frac{1}{4 \pi} \int_{\partial\Sigma} \mathrm{d}\xi - 2 \Lambda (*\omega) =  n\left(k-\frac{4 n^2}{l^2}\right),
\end{equation*}
where $k$ is the curvature index , $k=0$ for the flat-horizon, and $k=\pm 1$ for the spherical and hyperbolic horizons, respectively. Note that for regular thermodynamics where the AdS radius, $l$, is fixed, it follows that the NUT parameter $n$ is also a conserved quantity.\\
 \hfill\\
But what does the NUT parameter on the boundary correspond to? In the early works \cite{Chamblin:1998pz,Emparan:1999pm,Mann:1999pc}, where the authors used Euclidean signature, the NUT parameter "n" was interpreted as the squashing parameter controlling the deformation of the boundary manifold, i.e., the 3-sphere. A deeper interpretation appeared after studying the Taub-NUT solutions with Lorentzian signatures in the context of fluid/gravity correspondence\cite{Leigh:2011au, Caldarelli:2011idw}. It provided a link between the bulk properties and the hydrodynamics of its boundary quantum field theory. The NUT parameter was interpreted as a measure of vorticity in the flux of the dual conformal fluid \cite{Leigh:2011au}. It showed that a Taub-NUT-AdS solution is dual to a fluid characterized by a constant vortex flow \cite{Leigh:2011au, Caldarelli:2011idw} with a Dirac delta source, generated by the bulk Misner string. \\
\hfill\\
Many authors studied Taub-NUT solutions with Euclidean signature while enforcing the Misner periodicity condition \cite{Chamblin:1998pz,Mann:1999pc,Emparan:1999pm, Clarkson:2002uj,Astefanesei:2004kn}. They found that the thermodynamics of these solutions revealed unusual features. For example, the entropy is not the area of the horizon, even not always positive, and the horizon temperature and radius are fixed by the nut parameter n. Furthermore, in the $n\rightarrow 0$ limit, one expects to retrieve the known Schwarzschild-AdS temperature, instead, one gets a diverging temperature! Another unusual property of these solutions arises from the observation that the NUT charge, being a conserved quantity, should contribute an additional work term in the first law. However, no such contribution is present due to the relation between the nut charge and the horizon radius.\\
\hfill\\
Motivated by the above issues and some recent developments several authors \cite{Bordo:2019tyh, Bordo:2019slw, Chen:2019uhp, Hennigar:2019ive, BallonBordo:2019vrn, Wu:2019pzr, BallonBordo:2020mcs, Durka:2019ajz, Frodden:2021ces, Awad:2020dhy, Mann:2020wad, Abbasvandi:2021nyv, Awad:2022jgn, Awad:2023lyt, Tharwat:2023vku} tried to probe the possibility of relaxing the periodicity condition, leaving the Misner string visible and allowing the NUT charge to appear in the first law as an independent charge  (Please check the "Introduction" section of \cite{Awad:2022jgn} for a quick review). This new approach was first adopted by Bonner in \cite{Bonnor:1969ala} where he showed that the string singularity is a source of angular momentum. Also in \cite{Awad:2022jgn}, the authors found mass, angular momentum, NUT, electric, and magnetic charge densities distributed along Misner string when studying the dyonic Taub-NUT black hole solution. Moreover, other authors showed in \cite{Miller:1971em, Clement:2015cxa, Clement:2015aka} that the spacetime can be maximally extended and geodesically complete by relaxing the Misner periodicity condition without any causal pathologies.\\
\hfill\\
In this work we study the thermodynamics, with fixed cosmological constant, of the dyonic black branes with a non-vanishing NUT charge. For Taub–NUT solutions with flat horizons, the absence of a Misner string removes the need to impose the Misner periodicity condition \cite{Chamblin:1998pz}. Consequently, the NUT charge remains as an independent conserved quantity, contributing to the first law with an additional work term. Our work will closely follow the approach outlined in \cite{Awad:2022jgn,Awad:2023lyt}, but with two major differences. Firstly, instead of focusing on black holes with spherical horizons, we will be examining black holes with flat horizons, also known as "black branes". Secondly, while the authors in \cite{Awad:2023lyt} studied the extended thermodynamics of black holes, i.e., considers varying the cosmological constant as a thermodynamic variable, our study will focus on regular thermodynamics of black branes.
We show that allowing the dyonic black brane to have a non-vanishing NUT charge while relaxing the Misner periodicity condition results in a consistent thermodynamics where the Gibbs-Duhem relation and the first law are both satisfied, the entropy is related to the horizon area, and the temperature goes to that of the NUT-less case as $n \rightarrow 0$. We also show that the phase diagram becomes nontrivial in this case, with first-order phase transitions ending at critical points.\\
\hfill\\
The paper is organized as follows; in section (\ref{sec2}) we revisit the NUT-less dyonic black brane showing that its phase diagram is indeed trivial with no signs of phase transitions or critical behavior. In section (\ref{sec3}) we study the Taub-NUT dyonic black brane, studying the regular thermodynamics of the mixed ensemble of the solution by calculating its thermodynamic quantities and showing that they satisfy the first law and the Gibbs-Duhem relation in subsection (\ref{sec3.1}), In (\ref{sec3.2}) we divide the general case into sub-cases depending on the values of the thermodynamic quantities and studying the phase transitions and the phase diagram for each case. Finally in section (\ref{sec4}) we discuss our treatment and present our conclusion while outlining possible extensions and future directions.

	\section{NUT-less Dyonic AdS Black Brane}\label{sec2}
	The metric of the dyonic black brane in AdS space is given by
	\begin{equation}
		ds^2=-f(r)\,dt^2\,+\,\frac{dr^2}{f(r)}\,+\,\frac{r^2}{l^2}\,(\,dx^2+\,x^2\,d\varphi^2), \label{nutlessmetric}
	\end{equation}
	here $x \in [0, A]$, $\varphi \in [0, 2\pi]$, and $l$ is the anti-de-Sitter radius given by $l=\sqrt{-\frac{3}{\Lambda}}$ where $\Lambda$ is the cosmological constant. The area of the brane is given by 
	\begin{equation}
		\int_0^{2\pi}\int_{0}^{A} \frac{1}{l^2}\,x \, dx\,d\phi = \frac{\pi A^2}{l^2} = 4 \pi \sigma. \label{sigma-nutless}
	\end{equation}
	And $f(r)$ is given by
	\begin{equation}
		f(r)\,=\,\frac{r^2}{l^2} \, + \,\frac{q^2 \, + p^2}{r^2} \, - \,\frac{2\,m}{r} \,. \label{fr-nutless}
	\end{equation}
	Here $m$, $q$, and $p$ are constants, but as we will show in a short while they are related to the mass and charge densities of the brane.\\
	\hfill\\
	From the horizon condition, $f(r_h) = 0$, the mass parameter, $m$, of the brane is given by
	\begin{equation} \label{mass-nutless}
		m = \frac{1}{2} \left( \frac{r_h^3}{l^2} + \frac{q^2+p^2}{r_h}\right),
	\end{equation}
	The non-vanishing components of the gauge potential $A_\mu$ are given by
	\begin{equation} \label{At-nutless}
		A_t\, = \,V - \frac{q}{r},
	\end{equation}
	and
	\begin{equation} \label{Ap-nutless}
		A_{\varphi} \, = \, \frac{p \, x^2}{2 \, l^2} .
	\end{equation}
	The square of the gauge potential, $A_{\mu}A^\mu$, is given by
	\begin{equation} \label{A-sqr-nutless}
		A_{\mu}A^\mu\,= \,\frac{p^2\,x^2}{4\,r^2\,l^2}\, -  \,\frac{(r\,V\,-\,q)^2}{r^2\,f(r)}.
	\end{equation}
	To prevent $A_{\mu}A^\mu$ from blowing up at the horizon (see \cite{Hawking:1995ap,Gibbons:1976ue}), $q$  must be related to $V$ in the following way 
	\begin{equation} \label{q-phi-nutless}
		q\,=\, r_h \,V,
	\end{equation}
	The electric and magnetic potentials can be calculated as follows
	\begin{equation} \label{q-potcalc-nutless}
		\phi_e\,=\, A_{t} |_{\infty} \, - \, A_{t} |_{r_{h}} \, = V = \frac{q}{r_h},
	\end{equation}\begin{equation} \label{p-potcalc-nutless}
		\phi_m\,=\, B_{t} |_{\infty} \, - \, B_{t} |_{r_{h}}  = \frac{p}{r_h}.
	\end{equation}
	Where the one form $B$ is the solution for $dB\,=\,*F$, where $*F$ is the Hodge dual of $F$, given by
	\begin{equation} \label{F-Dual}
		(*F)_{\mu \nu} = \frac{1}{2}\, \epsilon^{\,\alpha \beta}_{\:\:\,\:\:\: \mu \nu} \, F_{\alpha \beta},
	\end{equation}
	where $\epsilon$ is the Levi-Civita tensor.	The electric and magnetic charge densities are calculated using Stokes' theorem.
	\begin{equation} \label{Qe-nutless}
		Q_e = \frac{1}{4 \pi \sigma} \int_{\partial\Sigma} *F \,  =  \, q,
	\end{equation}
	\begin{equation} \label{Qm-nutless}
		Q_m = - \frac{1}{4 \pi \sigma}	 \int_{\partial\Sigma} F \,=\, p.
	\end{equation}
	We must be careful when dealing with the boundary surface $\partial\Sigma$, as it is not simply the surface of constant $r$, but also consists of the surface of constant $x$.
	\begin{figure}[htp]
		\centering
		\includegraphics[width=0.7 \textwidth]{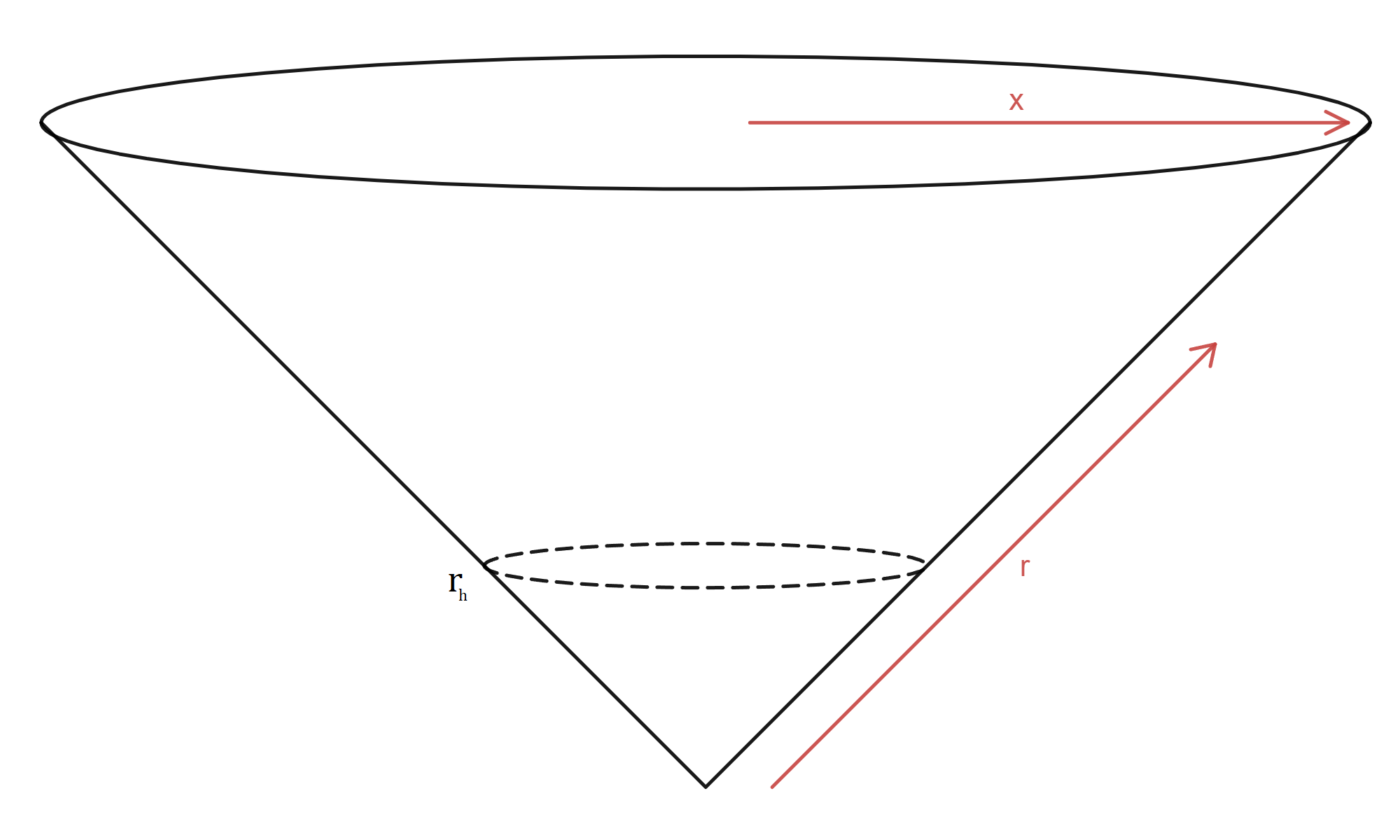}
		\caption{\footnotesize The boundary of spacetime at spatial infinity. the boundary consists of the top cape of $r=\infty$ and the side of the cone at $x = A$ from $r = r_h$ to $r = \infty$. The dashed line represents the hypersurface of $r = r_h$.}
		\label {boundry}
	\end{figure}
	\\\hfill
	From Eq.(\ref{Qe-nutless}) and Eq.(\ref{Qm-nutless}) we can see that $q$ and $p$ are proportional to the brane's electric and magnetic charges per unit area, $\sigma$, respectively.	
	\subsection{Thermodynamics}\label{sec2.1}
\begin{figure}[htp]
		\centering
		\includegraphics[width=0.8 \textwidth]{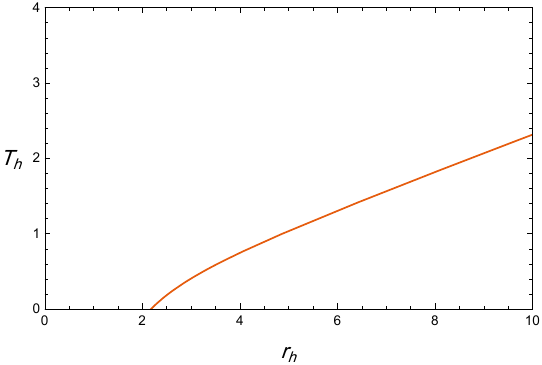}	\caption{\footnotesize The temperature of the dyonic black brane as a function of its horizon radius for $Q_m = 5$ and $\phi_e = 3$. The graph shows that for any temperature, there exists only one possible black brane solution.} 
		\label {T-r nutless}
	\end{figure}

\begin{figure}[htp]
		\centering
		\includegraphics[width=0.8 \textwidth]{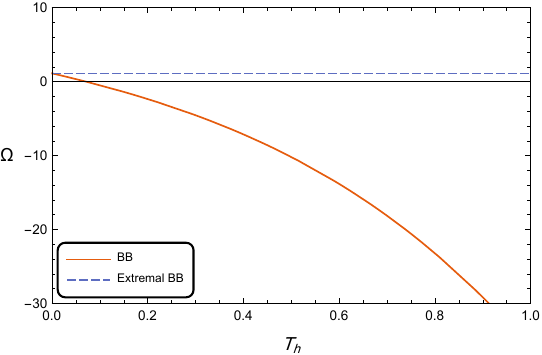}
		\caption{\footnotesize The grand potential densities of the possible phases as functions of their horizon temperature for $Q_m = 5$ and $\phi_e = 3$, the solid line represents the dyonic black brane, while the dotted line represents the extremal dyonic black brane. The graph shows that for any temperature the black brane solution always has a lower grand potential than the extremal black brane.} 
		\label {G-r nutless}
	\end{figure}

	The Euclidean path integral boundary conditions fix the boundary metric and the spatial component of the gauge potential, which in turn fixes the electric potential and the magnetic charge density. Therefore, the ensemble in hand is a mixed ensemble with the partition function $Z=Z(T,\phi_e, Q_m)$ \cite{Gibbons:1976ue, Hawking:1995ap}.\\
	\hfill\\
	The temperature of the horizon is calculated by performing a Wick rotation to the metric, with the temperature equal to the inverse of the periodicity of the imaginary time coordinate \cite{Hartle:1976tp, Gibbons:1976pt, Gibbons:1976ue}\\
	\begin{equation} \label{T-nutless}
		T_h \,=\, \frac{f'(r)}{4\,\pi} \biggr|_{r=r_h}\,=\, \frac{1}{\,4 \, \pi \, r_h} \left(\frac{3\,r_h^2}{l^2}  - \frac{Q_m^2}{r_h^2} -\phi_e^2 \right).
	\end{equation}
	From Eq.(\ref{T-nutless}) we can see that regardless of the values of $Q_m$ and $\phi_e$, the temperature is always a monotonic function in the horizon radius. This can be seen in Fig. \ref{T-r nutless}. 
	Moreover, we can see that there is always an extremal solution with radius $r_e$, where
	\begin{equation} \label{re-nutless}
		r_e = \frac{\sqrt{6 l^2\,\phi_e^2\,+\,6\,\sqrt{l^4\,\phi_e^4\,+\,12 l^2 Q_m^2}}}{6 }.
	\end{equation}
	As $f'(r)\Bigr|_{r=r_e} = 0$, the solution will not develop any conical singularity when the metric is Wick rotated. Therefore, the solution can have any arbitrary temperature, meaning it can exist at any temperature in that regard. However, as the surface gravity of the extremal black branes also vanishes, we do not expect that the extremal black brane can really exist at any temperature other than $T = 0$. Indeed, we will find that at any non-zero temperature, there will always exist another solution with a lower grand potential density.\\
	\hfill\\
	To calculate the grand potential density, one must first calculate the on-shell gravitational action. We will be calculating the on-shell action of the solution using the counter-term method \cite{Emparan:1999pm} such that
	\begin{equation} \label{action-nutless}
		I = I_b + I_s + I_{ct},
	\end{equation}
	where $I_b$ is the Einstein-Hilbert action with negative cosmological constant and the electromagnetic
	contribution to the action which is given by
	\begin{equation} \label{Ib-nutless}
		I_b = - \frac{1}{16\pi} \int_\mathcal{M} d^4x\,\sqrt{-g}\,(R-2\Lambda-F_{\mu\nu} F^{\mu\nu}),
	\end{equation}
	and $I_s$ is the Gibbons–Hawking–York boundary term given by
	\begin{equation} \label{Is-nutless}
		I_s = - \frac{1}{8\pi} \int_{\partial \mathcal{M}} d^3x\,\sqrt{-h}\,K,
	\end{equation}
	where $h$ is the determinant of the boundary metric and $K$ is the trace of the
	boundary's extrinsic curvature $K^{\alpha \beta}$. The integral $I_{ct}$ is the counter-term to cancel the divergences that arise from integrating over an infinite AdS volume, which is given by
	\begin{equation} \label{Ict-nutless}
		I_{ct} =  \frac{1}{8\pi} \int_{\partial \mathcal{M}} d^3x\,\sqrt{-h}\,\left(\frac{2}{l}+\frac{l}{2}\mathcal{R}\right),
	\end{equation}
	where $\mathcal{R}$ is the Ricci scalar for the boundary metric. The on-shell action density is evaluated to give
	\begin{equation} \label{action-calced-nutless}
		\mathcal{J} = \frac{I}{\sigma} = \beta \, \Omega = \frac{\beta}{2}\left(m+\frac{p^2-q^2}{ r_h}-\frac{r_h^3}{l^2}\right) =  \frac{\beta}{4}\left(\frac{3\,p^2}{r_h}-\frac{r_h^3}{l^2}-\phi_e^2 r_h\right).
	\end{equation}
	Where $\,\Omega\,$  is the grand potential density, which equals $\mathcal{J}/\beta$. The entropy density of the system can be calculated from the on-shell action density as follows
	\begin{equation} \label{entropy-nutless}
		S = \beta\,\partial_{\beta}\mathcal{J}-\mathcal{J} =  \pi r_h^2,
	\end{equation}
	which is consistent with the entropy being one-quarter of the black brane area. Moreover, the variation of $\,\Omega\,$ is consistent
	with Gibbs-Duhem relation
	\begin{equation} \label{Gibbs-Duhem-nutless}
		\mathrm{d}\Omega\,=\,-S\,\mathrm{d}T\,-\,Q_e\,\mathrm{d}\phi_e\,+\,\phi_m\,\mathrm{d}Q_m,
	\end{equation}
	with
	\begin{equation} \label{Gibbs-Duhem-nutlessI}
		\left(\frac{\partial{\Omega}}{\partial{T}}\right)_{\phi_e,Q_m} = -S , \quad \left(\frac{\partial{\Omega}}{\partial{\phi_e}}\right)_{T,Q_m} = -Q_e , \quad \left(\frac{\partial{\Omega}}{\partial{Q_m}}\right)_{T ,\phi_e} = \phi_m.
	\end{equation}
	Also, the total energy density of the solution can be calculated from the on-shell action density as follows
	\begin{equation} \label{total-energy-nutless}
		\mathfrak{M} = \partial_{\beta}\mathcal{J}\,+\,Q_e\,\phi_e = \frac{1}{2} \left(\frac{r_h^3}{l^2} + \frac{Q_m^2}{r_h}+\phi_e^2 \, r_h\right) =  m,
	\end{equation}
	which means that $m$ is also proportional to the mass per unit area of the brane. 
	The total energy density of the solution is also consistent with the first law
	\begin{equation} \label{first-law-nutless}
		\mathrm{d}\mathfrak{M}=T\mathrm{d}S+\phi_e\,\mathrm{d}Q_e+\phi_m\,\mathrm{d}Q_m,
	\end{equation}
	with
	\begin{equation} \label{first-law-nutlessII}
		\left(\frac{\partial{\mathfrak{M}}}{\partial{S}}\right)_{Q_e,Q_m} = T , \quad
		\left(\frac{\partial{\mathfrak{M}}}{\partial{Q_e}} \right)_{S,Q_m} = \phi_e, \quad \left(\frac{\partial{\mathfrak{M}}}{\partial{Q_m}} \right)_{S,Q_e} = \phi_m.
	\end{equation}\\
	Next, we need to check the thermal stability of the black brane. 
	We can see that the slope of the temperature curve is positive everywhere, Fig. \ref{T-r nutless}, indicating that the black brane is thermally stable at any temperature.\\
	\hfill\\
	To see that the dyonic black brane solution is always more thermodynamically preferable than its extremal solution as we mentioned earlier, the grand potential densities of both solutions are plotted in Fig. \ref{G-r nutless}. it is clear from the Fig. that the dyonic black brane solution is always the most thermodynamically preferable solution for any non-zero value of the horizon temperature.\\
	\hfill\\
	We can clearly see that for NUT-less dyonic black branes, the phase diagram is always trivial and phase transitions do not occur. In the next section, we will investigate the solution where the NUT charge is not zero to check if this will render a non-trivial phase structure.		
	\section{Taub-NUT Dyonic AdS Black Brane} \label{sec3}
	Now we will take a look at the same solution but with a non-vanishing NUT charge. More crucially, as we mentioned in the introduction, we will follow the same approach as in \cite{Awad:2023lyt}, where we relax the periodicity condition for the time coordinate, making the NUT charge a true independent charge that appears in the first law alongside its conjugate quantity $\phi_n$.\\
	\hfill\\
	The metric for the Taub-NUT dyonic AdS black brane is given by
	\begin{equation}
		ds^2=-f(r)\,\left( dt\,-\, \frac{n x^2}{l^2}\,d\varphi\right)^2 +\,\frac{dr^2}{f(r)}\,+\,\frac{r^2+n^2}{l^2}\,(\,dx^2+\,x^2\,d\varphi^2), \label{nutfulmetric}
	\end{equation}
	where the area of the brane and the ranges of both $x$ and $\phi$ are the same as those in the NUT-less case. The function $f(r)$ is now given by
	\begin{equation}  \label{fr-nutful}
    		f(r)\,=\,\frac{q^2+p^2-2\,m\,r}{r^2+n^2} \, + \, \frac{r^4+6\,n^2\,r^2-3\,n^4}{\left(r^2+n^2\right)\,l^2}.
	\end{equation}
	Again, $m$, $q$, and $p$ are constants. The proper mass and charge densities will be calculated later.\\
	\hfill \\
	Using $f(r_h)=0$, the mass parameter, $m$, is related to the other parameters by
	\begin{equation}  \label{m-nutful}
		m=\frac{l^2 \left(q^2+p^2\right) \, + \, r_h^4+6\,n^2\,r_h^2-3\,n^4}{2 l^2 r_h}.
	\end{equation}
	The non-vanishing components of the gauge potential $A_\mu$ are
	\begin{equation}  \label{At-nutful}
		A_t=\frac{pn-qr}{r^2+n^2}+V,
	\end{equation}
	and
	\begin{equation}  \label{Ap-nutful}
		A_\varphi=\frac{p\left(r^2-n^2\right)+2qnr}{2\,l^2\, \left(r^2+n^2\right)}\,x^2,
	\end{equation}
	To guarantee that the square of the gauge potential, $A_\mu A^\mu$, 
	is regular at the horizon, the parameters $q$, $n$, and $V$ 
	must be related through the relation
	\begin{equation} \label{q-phi-nutlful}
		q\,=\, \frac{pn+V\left(r_h^2+n^2\right) }{r_h},
	\end{equation} 
	The electric and magnetic potentials are calculated from the relations
	\begin{equation} \label{q-potcalc-nutful}
		\phi_e\,=\, A_{t} |_{\infty} \, - \, A_{t} |_{r_{h}} \, = V
	\end{equation}\begin{equation} \label{p-potcalc-nutful}
		\phi_m\,=\, B_{t} |_{\infty} \, - \, B_{t} |_{r_{h}}  = \frac{p+nV}{r_h}.
	\end{equation}
	The electric and magnetic charge densities are calculated using Stokes' theorem
	\begin{equation}\label{Qe-Qm-nutlfull}
		Q_e = \frac{1}{4 \pi \sigma} \int_{\partial\Sigma} *F, \quad \, Q_m = - \frac{1}{4 \pi \sigma}	 \int_{\partial\Sigma} F  
	\end{equation}
	As seen before, the hypersurface at spatial infinity consists of two parts; the surface of constant $r$ with $r=\infty$ and $x \in [0, A]$, and the surface of constant $x$ with $x = A$ and $r \in [r_h, \infty]$. The electric and magnetic charge densities are then given by
	\begin{equation*}\label{Qe-nutlfull}
		Q_e =\frac{1}{4 \pi \sigma} \int_{0}^{2\pi}\int_{0}^{A} (*F)_{x\phi} \, \mathrm{d}x \mathrm{d}\phi \, + \frac{1}{4 \pi \sigma}\int_{r_h}^{\infty}\int_{0}^{2\pi}(*F)_{\phi r} \, \mathrm{d}\phi \mathrm{d}r
	\end{equation*}
	\begin{equation}\label{QeII-nutlfull}
		= q - 2 n \phi_m,
	\end{equation}
	and,
	\begin{equation*}\label{Qm-nutlfull}
		Q_m =-\frac{1}{4 \pi \sigma} \int_{0}^{2\pi}\int_{0}^{A} (F)_{x\phi} \, \mathrm{d}x \mathrm{d}\phi \, - \frac{1}{4 \pi \sigma}\int_{r_h}^{\infty}\int_{0}^{2\pi}(F)_{\phi r} \, \mathrm{d}\phi \mathrm{d}r
	\end{equation*}
	\begin{equation}\label{QmII-nutlfull}
		= p + 2 n \phi_e.
	\end{equation}
	In fact, if we choose the boundary at any generic radial distance with $r \, > \, r_h$, we will get the same results for the electric and magnetic charge densities.
	This is a very interesting result. Unlike the NUT-less case where there was no flux through the constant $x$ hypersurface, we get here an $r$-dependent electric and magnetic charge densities if we only considered the surfaces of constant $r$ for any generic radial distance with $r > r_h$. This is a consequence of the flux leaking through the constant $x$ surface.
 
	\hfill\\
	We can also calculate the conserved NUT charge density, $N$, as the dual to the mass density given by the generalized Komar integral \cite{Kastor:2009wy, Bordo:2019tyh} 
	\begin{equation}\label{con-N}
		N =  \frac{1}{4 \pi \sigma} \int_{\partial\Sigma} \mathrm{d}\xi - 2 \Lambda (*\omega) =  \frac{4 n^3}{l^2},
	\end{equation}
	where $\xi = \partial_{t}$ is a time-like Killing vector and $\omega$ is the Killing potential satisfies $\xi^\mu = \nabla_\nu \omega^{\nu \mu}$  or $d\omega = *\xi $.

\subsection{Thermodynamics}\label{sec3.1}
	As discussed previously in the NUT-less case, the Euclidean path integral boundary conditions fix the electric potential and the magnetic charge density. In the NUTty case, they in addition fix the NUT charge density, $N$.  The ensemble therefore is also a mixed ensemble with the partition function $Z=Z(T,\phi_e, Q_m, N)$ \cite{Awad:2020dhy}.\\
	\hfill\\
	The temperature of the brane is given by
	\begin{equation} \label{n-Tutful}
		T_h \,=\, \frac{f'(r)}{4\,\pi} \,=\frac{\left(3r_h^2-l^2 \phi_e^2\right)\left(r_h^{2}+n^{2}\right)-l^2p\left(p+2n\phi_e\right)}{4\pi l^2 r^3_h}.
	\end{equation}
	As in the NUT-less case, we calculate the on-shell action density using the counter-term method, which yields
	\begin{equation} \label{action-calced-nutful}
		\mathcal{J} = \frac{I}{\sigma} = \beta \, \Omega =\frac{\beta }{2}\left(m-q\phi_e+\frac{\left(p+n\phi_e\right)\left(p+2n\phi_e\right)}{r_h}-\frac{r_h\left(r_h^2+3n^2\right)}{l^2}\right).
	\end{equation}
	The entropy density of the system is given by
	\begin{equation} \label{entropy-nutful}
		S = \beta\,\partial_{\beta}\mathcal{J}-\mathcal{J} = \pi \left(r_h^2+n^2\right),
	\end{equation}
	which is consistent with being one-quarter of the brane area.\\
	\hfil\\
	Earlier we calculated the conserved NUT charge density to be $\, N = \frac{4 n^3}{l^2}$. Given that we are doing a regular treatment to the thermodynamics, i.e. the cosmological constant is not allowed to vary and $l$ is constant, $n$ is consequently a conserved quantity. For the convenience of calculations, from now on we will continue our treatment using the conserved quantity $n$.\\
	\hfill \\
	It is important to see that the charge densities calculated earlier, $Q_e= \left(q-2n\phi_m\right)$ and $Q_m\,=\,\left(p\,+\,2n\phi_e\right)$, do not satisfy the first law and the Gibbs-Duhem relation. In fact, if we set the electric charge density to be $Q_e$, the magnetic charge density that satisfies the first law and the Gibbs-Duhem relation is $p$. And if we set the magnetic charge density to $Q_m$, the electric charge that satisfies the two relations is nothing but $q$. In short, to satisfy these relations one of the charges must be the charge at infinity and the other one the charge at the horizon! This fact was previously reported by several authors \cite{Bordo:2019slw,Chen:2019uhp,BallonBordo:2020mcs,Awad:2020dhy,Awad:2022jgn,Awad:2023lyt,Tharwat:2023vku}. Two of these authors \cite{Chen:2019uhp,Awad:2022jgn} showed that there exist an infinite family of electric and magnetic charge densities that satisfy both relations. The two charge densities can be parameterized by a real parameter $\alpha$, in the following forms:
	\begin{equation} \label{QQ-nutful}
		\widetilde{Q}_e(\alpha) =  q+\left(\alpha - 1 \right) n \phi_m  ,
	\end{equation}
	\begin{equation} \label{QQ-nutfulII}
		\widetilde{Q}_m(\alpha)=p+(\alpha+1) n \phi_e .
	\end{equation}
    Here we follow the same approach adopted in \cite{Awad:2022jgn, Awad:2023lyt} where the authors applied it to the dyonic Taub-NUT black holes with spherical horizons. However, in the case of the spherical horizon, $\alpha$ is fixed to some value that leaves the gauge potential regular along the Misner string. In our case, with a flat horizon, the lack of a Misner string allows $\alpha$ to become a free parameter. Later, we will see how the different choices of $\alpha$ affect the phase diagram of the solution. For now, the charges given by Eq.(\ref{QQ-nutful}) and Eq.(\ref{QQ-nutfulII}) are consistent with the Gibbs-Duhem relation
	\begin{equation} \label{Gibbs-Duhem-nutful}
		\mathrm{d}\Omega=-S\mathrm{d}T-\widetilde{Q}_e\mathrm{d}\phi_e+\phi_m\mathrm{d}\widetilde{Q}_m+\phi_n\mathrm{d}n
	\end{equation}
	with
	\begin{equation*}
		\left(\frac{\partial{\Omega}}{\partial{T}}\right)_{\phi_e,\widetilde{Q}_m ,n} = -S, \quad
		\left(\frac{\partial{\Omega}}{\partial{\phi_e}}\right)_{T,\widetilde{Q}_m,n} = -\widetilde{Q}_e,
	\end{equation*}
	\vspace{8pt}
	\begin{equation} \label{Gibbs-Duhem-nutfulII}
		\left(\frac{\partial{\Omega}}{\partial{\widetilde{Q}_m}}\right)_{T,\phi_e,n} = \phi_m, \quad \left( \frac{\partial{\Omega}}{\partial{n}}\right) _{T, \phi_e, \widetilde{Q}_m} = \phi_n,
	\end{equation}
	where $\phi_n$ is given by
	\begin{multline} \label{phi-n}
		\phi_n = \frac{1}{2\, n\,r_h}\bigg(\frac{3\,n^2- l^2\phi_e^2}{ l^2}\left( r_h^2-n^2\right)-r_h^2\,\phi_e^2\left(\alpha^2-1\right)
		\\
		+\frac{1}{\,r_h^2}\left(n\,\widetilde{Q}_m-\,\alpha\,\phi_e\left( r_h^2+n^2\right)\right)^2\bigg).
	\end{multline}
	Finally, the total energy density of the solution can be calculated from the on-shell action density as follows
	\begin{equation} \label{total-energy-nutful}
		\mathfrak{M} = \partial_{\beta}\mathcal{J}\,+\,\widetilde{Q}_e\,\phi_e,
	\end{equation}
	which evaluates to
	\begin{equation} \label{total-energy-nutfulII}
		\mathfrak{M} =\frac{ r_h\left( r_h^2+3n^2\right)}{2l^2}+\frac{\widetilde{Q}_m^2+\phi_e^2\left( r_h^2-\alpha^2 n^2\right)}{2r_h }=  m - n\phi_n.
	\end{equation}
	this satisfies the first law
	\begin{equation} \label{first-law-nutfull}
		\mathrm{d}\mathfrak{M}=T\mathrm{d}\,S+\phi_e\,\mathrm{d}\widetilde{Q}_e+\phi_m\,\mathrm{d}\widetilde{Q}_m + \phi_n\,\mathrm{d}n,
	\end{equation}
	with
	\begin{equation*} \label{first-law-nutfulII}
		\left(\frac{\partial{\mathfrak{M}}}{\partial{S}}\right)_{\widetilde{Q}_e,\widetilde{Q}_m,n} = T , \quad
		\left(\frac{\partial{\mathfrak{M}}}{\partial{\widetilde{Q}_e}} \right)_{S,\widetilde{Q}_m,n} = \phi_e,
	\end{equation*}
	\vspace{8pt}
	\begin{equation} \label{first-law-nutfulIII}
		\left(\frac{\partial{\mathfrak{M}}}{\partial{\widetilde{Q}_m}}\right)_{S,\widetilde{Q}_e,n} = \phi_m , \quad
		\left(\frac{\partial{\mathfrak{M}}}{\partial{n}} \right)_{S,\widetilde{Q}_e,\widetilde{Q}_m} = \phi_n.
	\end{equation}
	We notice from Eq.(\ref{total-energy-nutfulII}) that unlike the NUT-less case, where the total energy density is just $m$, there exists an extra term here, $n \phi_n$, which also contributes to the total energy density of the solution.\\
	\subsection{Phase Transitions and Critical Points}\label{sec3.2}
    The temperature written in terms of the independent thermodynamic quantities takes the form
	\begin{equation} \label{T-Q_m}
		T_h \,=\frac{3 \left( r_h^{2}+n^2\right)}{4\pi l^2 r_h}-\frac{  r_h^2 \phi_e^2 + \left(\widetilde{Q}_m -  \alpha n \phi_e \right)^2}{4\pi r^3_h}.
	\end{equation}
    Before starting our analysis, we can notice that the signs of $\alpha$, $\widetilde{Q}_m$, and $\phi_e$ only affect the term $\left(\widetilde{Q}_m -  \alpha n \phi_e \right)^2$. If two of the variables switch signs, the system is unaffected, but if all of them or only one them switches sign, we will get a system where $n \longrightarrow -n$. In the following analysis we will use positive values for $\alpha$, $\widetilde{Q}_m$, and $\phi_e$ knowing that we are not losing any generality.\\
    \hfill\\
    Since the Dyonic NUTless case has no phase transitions to compare it to the present cases, we find it convenient to construct the phase diagram as a $n-T$ diagram to highlight the role of the NUT charge in developing first order phase transitions and critical behaviors.\\
    \hfill\\
    By allowing non-zero values for $n$, we aim to check if we will get a non-monotonic behavior for $T_h(r_h)$, which indicates first order phases, and, if so, we require to get the critical points, which are the values of $n$ at which $T_h(r_h)$ goes from being monotonic to non-monotonic. Doing so, we got 4 critical points in general
     \begin{equation} \label{Nc-gI}
		n_{c_{1,2}}= -  \alpha l \phi_e \mp \frac{1}{3}{\sqrt{ 3 l  \left(l \phi_e^2\, (3\alpha^2+1) + 6 \widetilde{Q}_m \right)}},
	\end{equation}\\
    corresponding to
    \begin{equation} \label{rc12-gI}
		r_{h(c_{1, 2})} =\frac{1}{3}{\sqrt{9\left(\alpha^2l ^2 \phi_e^2+l\widetilde{Q}_m \right) \pm 3 l \sqrt{3\alpha^2 l \phi_e^2 \left( l \phi_e^2\,(3\alpha^2+1)+6 \widetilde{Q}_m\right)}}}\,
	\end{equation}\\
    and
    \begin{equation} \label{Nc-gII}
		n_{c_{3,4}}=   \alpha l \phi_e \mp \frac{1}{3}{\sqrt{ 3 l  \left(l \phi_e^2\, (3\alpha^2+1) - 6 \widetilde{Q}_m \right)}},
	\end{equation}\\
    corresponding to
    \begin{equation} \label{rc34-gI}
		r_{h(c_{3, 4})} =\frac{1}{3}{\sqrt{9\left(\alpha^2l ^2 \phi_e^2-l\widetilde{Q}_m \right) \mp 3 l \sqrt{3\alpha^2 l \phi_e^2 \left( l \phi_e^2\,(3\alpha^2+1)-6 \widetilde{Q}_m\right)}}},
	\end{equation}\\
    Some of these expressions may become imaginary, which correspond to non-physical critical points. However, for each critical point, the same square-root term that determine the reality of $n_c$ also appear in $r_{h(c)}$. Thus, ensuring the reality of $r_{h(c)}$ automatically guarantees that $n_c$ is also real.\\
    \hfill\\
    Assuming positive $\widetilde{Q}_m$ and $\phi_e$, one can see that while the first critical point is always physical, the other critical points come with restrictions on the thermodynamic quantities in order to physically exist.\\
    \hfill\\
    \begin{itemize}
        \item Second critical point:\\
        For this critical point to exist, relation for $r_{h(c_2)}$ imposes the following condition
        \begin{equation} \label{r2_con}
           \widetilde{Q}_m > \frac{\alpha l \phi^2}{\sqrt{3}}
        \end{equation}
        \item Third critical point:\\ 
       The conditions required for $r_h(c_3)$ to be real are
         \begin{equation} \label{r3_con1}
           \alpha > \frac{\sqrt{3}}{3} 
        \end{equation}
        and
        \begin{equation} \label{r3_con2}
         \frac{\alpha l \phi^2}{\sqrt{3}} < \widetilde{Q}_m  \le \frac{1}{6} \left[l \phi^2 \left(3 \alpha^2+1 \right) \right].
        \end{equation}
        \item Fourth critical point:\\
        Finally for the fourth critical point, $r_{h(c_4)}$ is real when
        \begin{equation} \label{r4_con2}
           \alpha \leq \frac{\sqrt{3}}{3}, \quad \text{and} \quad \widetilde{Q}_m  < \frac{\alpha l \phi^2}{\sqrt{3}},
        \end{equation}
        or
        \begin{equation} \label{r4_con1}
            \alpha > \frac{\sqrt{3}}{3}, \quad \text{and} \quad \widetilde{Q}_m  \le \frac{1}{6} \left[l \phi^2 \left(3 \alpha^2+1 \right) \right].
        \end{equation}
    \end{itemize}
    From the above conditions we can see that the third critical point is always non-physical for $\alpha \leq \frac{\sqrt{3}}{3}$ regardless of the values of $\widetilde{Q}_m$ and $\phi_e$. Also, for $\alpha \leq \frac{\sqrt{3}}{3}$, the second and fourth critical points cannot be real simultaneously, implying that at most only two critical points can exist simultaneously in this regime. On the other hand, for $\alpha > \frac{\sqrt{3}}{3}$, all four critical points can be real simultaneously depending on the values of $\widetilde{Q}_m$ and $\phi_e$. Accordingly, we will split our analysis depending on the value of $\alpha$. Dependencies on the other thermodynamic quantities will be considered as sub-cases. In our analysis we chose to fix the value of $\phi_e$ and parametrize the sub-cases using $\widetilde{Q}_m$. The reverse, fixing the value of $\widetilde{Q}_m$ and parametrize the sub-cases with $\phi_e$, is also valid and will lead to the same results. Therefore, this choice will not lead to any loss of generality.\\
	\hfil\\
	\subsubsection{Case I: $\alpha = 0$}
	For $\alpha = 0$, the $n-T$ phase diagram will only depend on the value of $\widetilde{Q}_m$. Generally, we have two cases;
	
	\subsubsection*{Case I.I: $\widetilde{Q}_m = 0$}
	When fixing the values of $\alpha$ and $\widetilde{Q}_m$ to zero, the horizon temperature will take the form
	\begin{equation} \label{T-Q_m-case_1_1}
		T_h \,=\frac{3 \left( r_h^{2}+n^2\right)}{4\pi l^2 r_h}-\frac{  \phi_e^2}{4\pi  r_h}.
	\end{equation}
	This case is very similar to the grand canonical ensemble of Reissner-Nordstr\"{o}m-anti-de Sitter (RNAdS) black hole \cite{Chamblin:1999hg, Chamblin:1999tk}.
	\begin{figure}[htp]
		\centering
		\includegraphics[width=0.8 \textwidth]{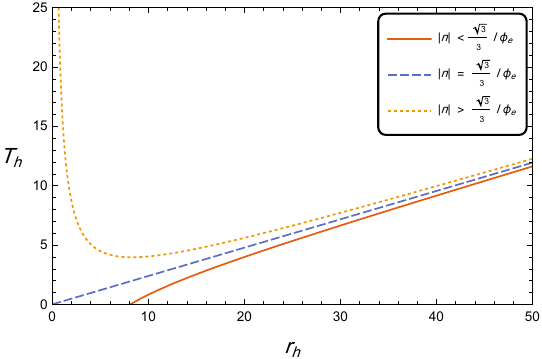}
		\caption{\footnotesize The black brane horizon temperature as a function of the horizon radius in case I.I for different values of $n$. The values used to draw this graph are ($l = 1$, $\phi_e = 15$, $\alpha = 0$, $\widetilde{Q}_m = 0$). The graph shows that for $n \leq \frac{\sqrt{3}}{3}l\phi_e$ the is only one possible black brane at any temperature, while for $n > \frac{\sqrt{3}}{3}l\phi_e$ there are two possible black brane phases for $T_h \geq T_{min}$ and no possible black brane phase for $T_h < T_{min}.$}
		\label {A0-Q0_T-r}
	\end{figure}
	\hfill\\
	As shown in Fig.(\ref{A0-Q0_T-r}), if $\abs{n} \leq \frac{\sqrt{3}}{3} l \phi_e$, the horizon temperature is a monotonic function of $r_h$, meaning that only one black brane can be in thermal equilibrium at any temperature. Also, there is always an extremal black brane with a horizon radius $r_e$ given by
	\begin{equation} \label{re-A0Q0}
		r_e = \frac{\sqrt{3 \left(l^2\,\phi_e^2-3n^2\right)}}{3}.
	\end{equation}
	However, if $\abs{n} > \frac{\sqrt{3}}{3} l \phi_e$, we see that the horizon temperature has a minimum, $T_{min}$, at $r_h = r_{min}$ where
	\begin{equation} \label{Q0-A0_r_min}
		r_{min}= \frac{\sqrt{3 \left(3 n^2- l^2 \phi_e^2\right)}}{3} ,
	\end{equation}
	and
	\begin{equation} \label{Q0-A0_T_min}
		T_{min} \,=\frac{\sqrt{9n^2-3 l^2 \phi_e^2}}{2\pi l^2}.
	\end{equation}
	It follows that at $T>T_{min}$ there are two possible black branes, a big black brane, and a small black brane, there is only one possible black brane at $T = T_{min}$, and finally, if $T < T_{min}$ no black brane can exist in thermal equilibrium, since the black brane horizon temperature cannot fall below $T_{min}$. Therefore, the black brane cannot be considered as a phase at this range of temperatures. \\
	\hfill\\
	The Schwarzschild AdS solution \cite{Hawking:1982dh} and some cases of the grand canonical ensemble of Reissner-Nordstr\"{o}m AdS solution \cite{Chamblin:1999hg, Chamblin:1999tk} also have a minimum value for $T_h$. However, these cases always have a pure thermal solution that exists at any temperature which shares the same asymptotic metric as the black hole. Our case lacks such a pure thermal solution. In fact, there is no known solution with the same asymptotic metric and have the same conserved charges as the Taub-NUT dyonic AdS black brane solution. This means that in our case there are no known solution that can exist at $T < T_{min}$. An educated guess would be that for $T < T_{min}$ the system will transit to a soliton-like solution, along the line that was presented in \cite{Surya:2001vj}. However, further investigation is required to confirm this assumption. \\
	\hfill\\
	Using the same arguments as in the NUT-less case, we can see that the black brane is thermally stable when $\left(\frac{\partial{T}}{\partial{r_h}}\right)_{\phi_e, \widetilde{Q}_m,n} \geq 0$. In summary:
	\begin{itemize}
		\item For $\abs{n} > \frac{\sqrt{3}}{3} l \phi_e$ and $T<T_{min}$, there are no possible black brane phases.
		\item For $\abs{n} > \frac{\sqrt{3}}{3} l \phi_e$ and $T=T_{min}$, the only existing phase is one black brane, and it is thermally stable.
		\item For $\abs{n} > \frac{\sqrt{3}}{3} l \phi_e$ and $T>T_{min}$, there are two possible black brane phases, the small black brane is thermally unstable while the big black brane is thermally stable.
		\item For $\abs{n} \leq \frac{\sqrt{3}}{3} l \phi_e$, the only existing phase is one black brane, and it is thermally stable.
	\end{itemize}
	\subsubsection*{Phase Diagram}
\begin{figure*}[h]
		\centering
		\begin{subfigure}[htp]{0.49\textwidth}   
			\centering 
			\includegraphics[width=1 \textwidth]{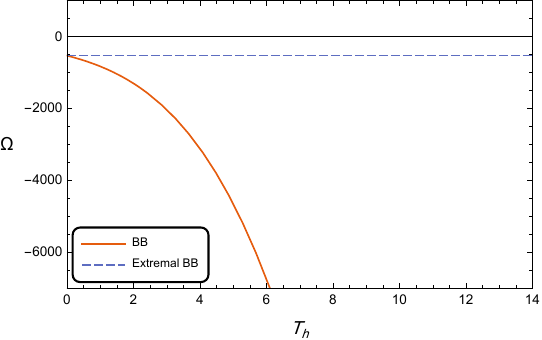}
			\caption[]%
			{{\small $\abs{n} < \frac{\sqrt{3}}{3} l \phi_e$}}    
		\end{subfigure}
		\hfill
		\begin{subfigure}[htp]{0.49\textwidth}   
			\centering \includegraphics[width=1 \textwidth]{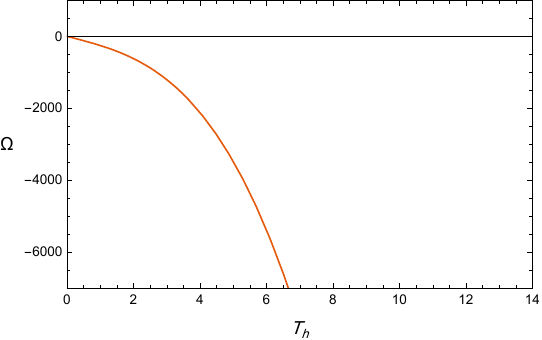}
			\caption[]%
			{{\small $\abs{n} = \frac{\sqrt{3}}{3} l \phi_e$}}    
		\end{subfigure}
		\vskip\baselineskip
		\begin{subfigure}[htp]{0.49\textwidth} 
			\centering \includegraphics[width=1 \textwidth]{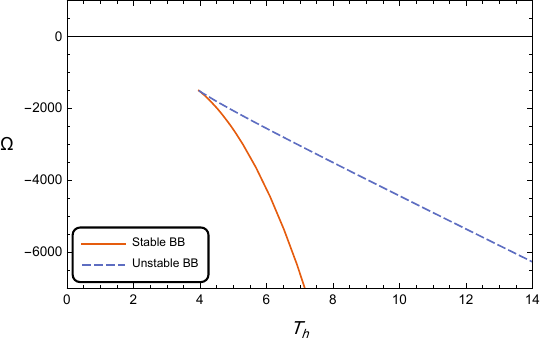}
			\caption[]%
			{{\small $\abs{n} > \frac{\sqrt{3}}{3} l \phi_e$}}    
		\end{subfigure}
		\caption{\footnotesize The grand potential density of the black brane as a function of the horizon temperature in case I.I for different values of $n$. The values used to draw this graph are ($l = 1$, $\phi_e = 15$, $\alpha = 0$, $\widetilde{Q}_m = 0$). The graphs show that for $n \leq \frac{\sqrt{3}}{3}l\phi_e$ the grand potential of the black brane is always lower than that of the extremal black brane, while for $n > \frac{\sqrt{3}}{3}l\phi_e$ there is no extremal black brane, and the stable black brane always have a lower grand potential than the unstable black brane.}
		\label{A0_Q0_G-T}
	\end{figure*}

\begin{figure}[htp]
		\centering
		\includegraphics[width=0.8 \textwidth]{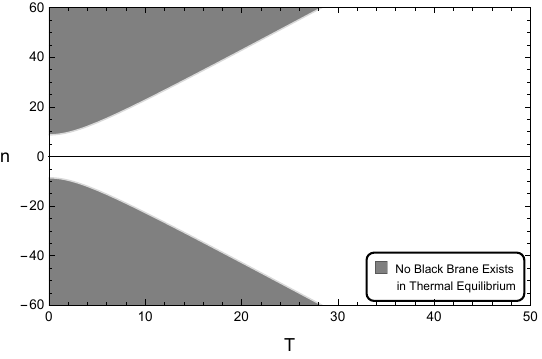}
		\caption{\footnotesize The $n-T$ phase diagram for case I.I. The values used to draw this graph are ($l = 1$, $\phi_e = 15$, $\alpha = 0$, $\widetilde{Q}_m = 0$). The graph shows that there exists only one phase everywhere except for the shaded gray area which marks where $T < T_{min}(n)$ and hence no black brane can exist in thermal equilibrium.}
		\label {A0-Q0_N-T}
	\end{figure}

	To further investigate the different cases and the phases they contain, the grand potential density of the solution as a function of its horizon temperature is plotted for different values of $n$. This is shown in Fig. \ref{A0_Q0_G-T}\\
	\hfill\\
	From Fig. \ref{A0_Q0_G-T} we can see that for $\abs{n} < \frac{\sqrt{3}}{3} l \phi_e$ the non-extremal black brane always has a lower potential density than the extremal one, making it the thermodynamically preferable phase. For $\abs{n} > \frac{\sqrt{3}}{3} l \phi_e$ we can see the absence of an extremal black brane. We can also see that no black brane exists with $T_h < T_{min}$. On the other hand, for $T > T_{min}$ two black branes exist, with the thermally stable one always having a lower potential density than the non-stable one.\\
	\hfill\\
	We are now fully acquainted with the required information to draw the $n-T$ phase diagram for this case. This is shown in Fig. \ref{A0-Q0_N-T}. We can see from the figure that there is only one stable black brane phase everywhere except for the gray-shaded area, in which $T < T_{min}(n)$, where black brane cannot exist as a phase.
	
	\subsubsection*{Case I.II: $\widetilde{Q}_m \neq  0$}
\begin{figure}[htp]
		\centering
		\includegraphics[width=0.8 \textwidth]{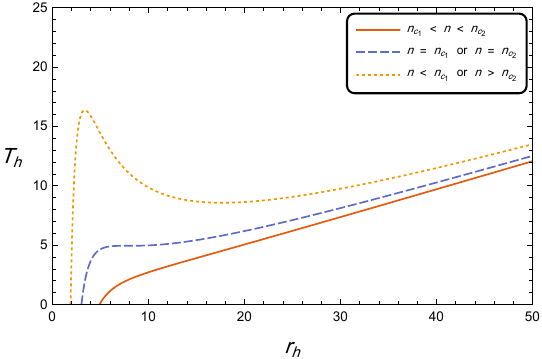}
		\caption{\footnotesize The black brane horizon temperature as a function of the horizon radius in case I.II for different values of $n$. Values used to draw this graph are ($l = 1$, $\phi_e = 15$, $\alpha = 0$, $\widetilde{Q}_m = 60$).
		}
		\label {A0-Q60_T-r}
	\end{figure}

	The non-vanishing magnetic charge density will make this case more interesting. The horizon temperature will take the form
	\begin{equation} \label{T-Q_m-case_1_2}
		T_h \,=\frac{3 \left( r_h^{2}+n^2\right)}{4\pi l^2 r_h}-\frac{\widetilde{Q}_m^2 + r_h^2\phi_e^2 }{4\pi  r_h^3}.
	\end{equation}\\
	Due to the non-vanishing $\widetilde{Q}_m$, an additional $-\frac{1}{r_h^3}$ term is present in the temperature formula. Fig. \ref{A0-Q60_T-r} shows that regardless the value of $n$, there exists an extremal black brane with horizon radius $r_e$ given by
	\begin{equation} \label{re-case_1_2}
		r_e = \frac{\sqrt{6 \left( l^2\,\phi_e^2-3n^2\right)\,+\,6\,\sqrt{\left( l^2\,\phi_e^2-3n^2\right)^2\,+\,12 l^2\widetilde{Q}_m^2}}}{6}.
	\end{equation}\\
	\hfill\\
	However, we either get a monotonic or non-monotonic $T_h(r_h)$ depending on the value of $n$. with two critical points
	\begin{equation} \label{Nc-case_1_2}
		n_{c_1}= - \frac{\sqrt{18 l \widetilde{Q}_m + 3 l^2 \phi_e^2}}{3}, \quad n_{c_2}=  \frac{\sqrt{18 l \widetilde{Q}_m + 3 l^2 \phi_e^2}}{3},
	\end{equation}\\
	both corresponding to
	\begin{equation} \label{rc-case_1_2}
		r_{h(c)} =\sqrt{l \widetilde{Q}_m},
	\end{equation}\\
	and
	\begin{equation} \label{Tc-case_1_2}
		T_c =\frac{2\sqrt{l \widetilde{Q}_m}}{\pi l^2}.
	\end{equation}\\
	For $n < n_{c_1}$ the horizon temperature of the black brane has a minimum value, $T_{min}$, at $r_h=r_{min}$ and a maximum value, $T_{max}$, at $r_h = r_{max}$. For $T < T_{min}$ or $T>T_{max}$ there exists only one black brane, while for $T \in (T_{min}, T_{max})$ there exist three black branes. By increasing the value of $n$ the intervals $(T_{min},T_{max})$ and $(r_{min},r_{max})$ shrink until $n = n_{c_1}$, where $T_{min}$ and $T_{max}$ converge at $T_{c}$ and also $r_{min}$ and $r_{max}$ converge at $r_{h(c)}$. At this point, two branches of the black brane solution merge, while the third one disappears.\\
	\hfill \\
	Increasing $n$ further, $T_h$ remains monotonic until $n = n_{c_2}$, after which the black brane solution re-splits into two branches while the third one reappears. The horizon temperature acquires minimum and maximum values again, with the intervals $(T_{min}, T_{max})$ and $(r_{min},r_{max})$ keep widening as $n$ continues to increase.\\
	\hfill\\
	From our previous discussion about the thermal stability of the black brane and from
	Fig. \ref{A0-Q60_T-r} we can then summarize the cases:
	\begin{itemize}
		\item For $n < n_{c_1}$ or $n > n_{c_2}$, and $T<T_{min}$ or $T>T_{max}$, the only existing phase is one black brane, and it is thermally stable.
		\item For $n < n_{c_1}$ or $n > n_{c_2}$, and $T \in (T_{min}, T_{max})$, there are three possible black brane phases, the small and the big black branes are thermally stable while the middle one is thermally unstable.
		\item For $n < n_{c_1}$ or $n > n_{c_2}$, and $T= T_{min}$ or $T=T_{max}$, there are two possible black brane phases, and both of them are thermally stable.
		\item For $n \in [n_{c_1}, n_{c_2}]$, the only existing phase is one black brane, and it is thermally stable.
	\end{itemize}
	\subsubsection*{Phase Diagram}
\begin{figure*}[h]
		\centering
		\begin{subfigure}[htp]{0.49\textwidth}   
			\centering 
			\includegraphics[width=1 \textwidth]{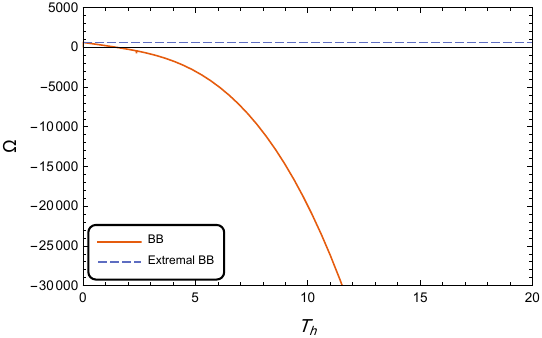}
			\caption[]%
			{{\small $n_{c_1} < n < n_{c_2}$}}    
		\end{subfigure}
		\hfill
		\begin{subfigure}[htp]{0.49\textwidth}   
			\centering 
			\includegraphics[width=1 \textwidth]{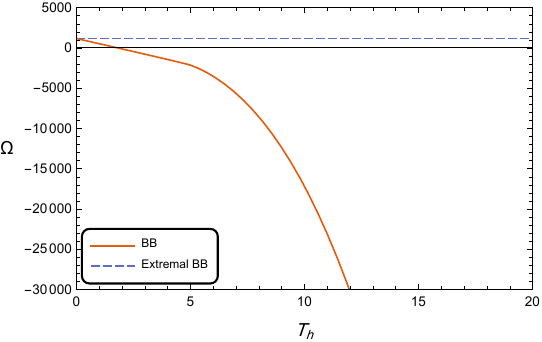}
			\caption[]%
			{{\small $n = n_{c_1}$ or $n = n_{c_2}$}}    
		\end{subfigure}
		\vskip\baselineskip
		\begin{subfigure}[htp]{0.49\textwidth} 
			\centering 
			\includegraphics[width=1 \textwidth]{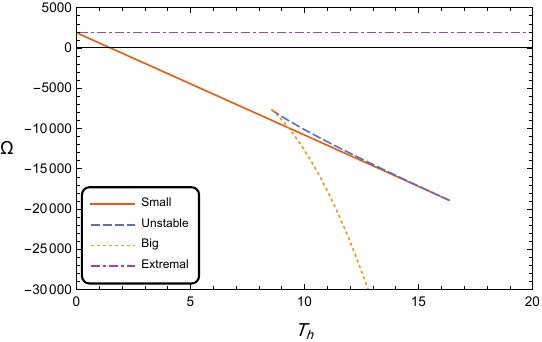}
			\caption[]%
			{{\small $n < n_{c_1}$ or $n > n_{c_2}$}}    
		\end{subfigure}
		\caption{\footnotesize The grand potential density of the black brane as a function of the horizon temperature in case I.II for different values of $n$. The values used to draw this graph are ($l = 1$, $\phi_e = 15$, $\alpha = 0$, $\widetilde{Q}_m = 60$).}
		\label{A0_Q60_G-T}
	\end{figure*}

	As before, we first plot the grand potential density of the solution as a function of the horizon temperature, as it contains all the necessary information to explore the phase diagram. This is shown in Fig. \ref{A0_Q60_G-T}. 
	\hfill \\
	\hfill\\
	From Fig. \ref{A0_Q60_G-T}, we can immediately notice the swallowtail behavior of the grand potential density for $n < n_{c_1}$ and $n > n_{c_2}$, signifying a first-order phase transition of the system. The grand potential densities of the small and the big black branes cross at $T = T_{tr}$, which marks a first-order phase transition between the two black branes. 
	Also, we can notice that the unstable middle black brane is never thermodynamically preferred. Moreover, for any value of $n$ and non-zero temperatures, there is always a phase with grand potential density lower than the extremal black brane, meaning that extremal black brane is also never thermodynamically preferred. Summarizing, we have
	\begin{itemize}
		\item For $n < n_{c_1}$ or $n > n_{c_2}$, and $T<T_{min}$ or $T>T_{max}$, the only existing phase is one black brane.
		\item For $n < n_{c_1}$ or $n > n_{c_2}$, and $T \in [T_{min}, T_{tr})$, there are two stable black brane phases, the small black brane is thermodynamically preferred while the big black brane is metastable.
		\item For $n < n_{c_1}$ or $n > n_{c_2}$, and $T= T_{tr}$, a first-order phase transition occurs between the small and the big black branes.
		\item For $n < n_{c_1}$ or $n > n_{c_2}$, and $T \in (T_{tr}, T_{max}]$, there are two stable black brane phases, the big black brane is thermodynamically preferred while the small black brane is metastable.
		\item For $n = n_{c_1}$ or $n = n_{c_2}$, and $T = T_c$, there is a critical point where the phase transition between the small and the big black branes vanishes and the two branes become indistinguishable.
		\item For $n \in [n_{c_1},n_{c_2}]$, the only existing phase is one black brane.
	\end{itemize}
	\begin{figure}[htp]
		\centering
		\includegraphics[width=0.8 \textwidth]{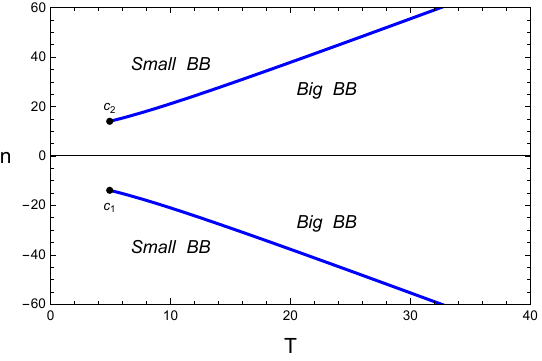}
		\caption{\footnotesize The $n-T$ phase diagram for case I.II. Values used to draw this graph are ($l = 1$, $\phi_e = 15$, $\alpha = 0$, $\widetilde{Q}_m = 60$).}
		\label {A0-Q60_N-T}
	\end{figure}
	\hfill\\
	From Fig. \ref{A0-Q60_N-T} and all the subsequent phase diagrams, we can see that no phase transitions or critical points exist along the line $n=0$ regardless of the values of the other thermodynamic quantities. This confirms that a non-zero NUT charge is the reason for these behaviors, as they disappear when $n$ vanishes. It is also important to stress that this is the first study to report rich phase structure with first-order phase transitions and critical behavior when studying the regular thermodynamics of black holes with flat horizons without adding extra terms to the gravitational action.
	
	\subsubsection{Case II: $\abs{\alpha} \in (0,\frac{\sqrt{3}}{3}]$}
	The non-vanishing $\alpha$ will make this case even more rich and understandably more complicated. With all of our thermodynamic quantities non-vanishing, the horizon temperature will take the general form in Eq.(\ref{T-Q_m}). Following the same approach and in light of our analysis of the critical point, we divide this case into sub-cases depending on the value of $\widetilde{Q}_m$.
	\subsubsection*{Case II.I: $\abs{\widetilde{Q}_m} < \frac{\alpha \, l \, \phi_e^2}{\sqrt{3}}$}\label{Case II.I}
\begin{figure}[htp]
		\centering
		\includegraphics[width=0.8 \textwidth]{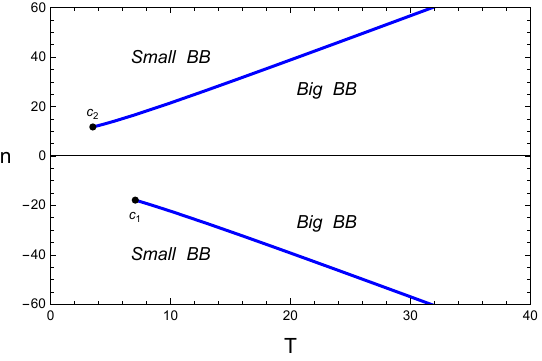}
		\caption{\footnotesize The $n-T$ phase diagram for case II.I. Values used to draw this graph are ($l = 1$, $\phi_e = 15$,$\alpha = \frac{\sqrt{3}}{5}$, $\widetilde{Q}_m = 30$).}
		\label {AII_Q30_N-T}
	\end{figure}

	This case is very close to case I.II in that regardless of the value of $n$ we always have an extremal solution with horizon radius $r_e$ given by
	\begin{equation} \label{re-general}
		r_e = \frac{\sqrt{6 \left( l^2\,\phi_e^2-3n^2\right)\,+\,6\,\sqrt{\left( l^2\,\phi_e^2-3n^2\right)^2\,+\,12 l^2\left( \alpha n \phi_e-\widetilde{Q}_m\right)^2}}}{6},
	\end{equation}\\
	and depending on the value of $n$ we either have monotonic or non-monotonic $T_h(r_h)$ with two critical values of $n$,
    \begin{equation*} \label{Nc-case_2_1I}
		n_{c_1}= -  \alpha l \phi_e - \frac{\sqrt{ 3 l  \left(l \phi_e^2\, (3\alpha^2+1) + 6 \widetilde{Q}_m \right)}}{3} ,
	\end{equation*}\\
     \begin{equation} \label{Nc-case_2_1II}
		n_{c_2}=  \alpha l \phi_e + \frac{\sqrt{ 3 l  \left(l \phi_e^2\, (3\alpha^2+1) - 6 \widetilde{Q}_m \right)}}{3}, 
	\end{equation}\\
	corresponding to
	\begin{equation*} \label{rc-case_2_1I}
		r_{h(c_1)} =\frac{\sqrt{9\left(\alpha^2l ^2 \phi_e^2+l\widetilde{Q}_m \right)+3 l \sqrt{3\alpha^2 l \phi_e^2 \left( l \phi_e^2\,(3\alpha^2+1)+6 \widetilde{Q}_m\right)}}}{3},
	\end{equation*}\\
	\begin{equation} \label{rc-case_2_1II}
		r_{h(c_2)} =\frac{\sqrt{9\left(\alpha^2l ^2 \phi_e^2-l\widetilde{Q}_m \right)+3 l \sqrt{3\alpha^2 l \phi_e^2 \left( l \phi_e^2\,(3\alpha^2+1)-6 \widetilde{Q}_m\right)}}}{3},
	\end{equation}\\
	and
	\begin{equation*} \label{Tc-case_2_1I}
		T_{c_1} =\frac{2}{3 \pi l^2} \sqrt{9\left(\alpha^2l ^2 \phi_e^2+l\widetilde{Q}_m \right)+3 l \sqrt{3\alpha^2 l \phi_e^2 \left( l \phi_e^2\,(3\alpha^2+1)+6 \widetilde{Q}_m\right)}},
	\end{equation*}\\
	\begin{equation} \label{Tc-case_2_1II}
		T_{c_2} =\frac{2}{3 \pi l^2}\sqrt{9\left(\alpha^2l ^2 \phi_e^2-l\widetilde{Q}_m \right)+3 l \sqrt{3\alpha^2 l \phi_e^2 \left( l \phi_e^2\,(3\alpha^2+1)-6 \widetilde{Q}_m\right)}}.
	\end{equation}
	\begin{figure}[htp]
		\centering
		\includegraphics[width=0.8 \textwidth]{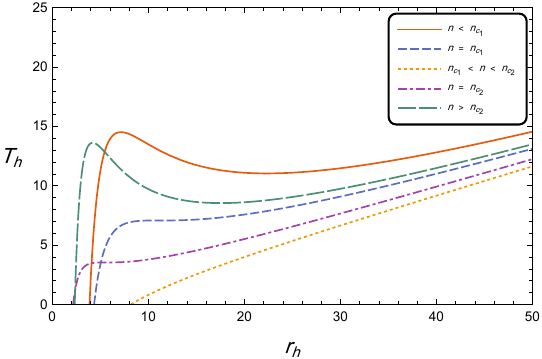}
		\caption{\footnotesize The black brane horizon temperature as a function of the horizon radius in case II.I for different values of $n$. Values used to draw this graph are ($l = 1$, $\phi_e = 15$,  $\alpha = \frac{\sqrt{3}}{5}$, $\widetilde{Q}_m = 30$).}
		\label {AII_Q30_T-r}
	\end{figure}
	\hfill\\
	However, as we can see from Fig. \ref{AII_Q30_T-r}, unlike case I.II, $n_{c_1} \neq -n_{c_2}$, $r_{h(c_1)} \neq r_{h(c_2)}$, and $T_{c_1} \neq T_{c_2}$, except for the special case when $\widetilde{Q}_m = 0$, giving rise to a non-symmetric phase diagram under $n \longrightarrow -n$ in general, Fig. \ref{AII_Q30_N-T}, except for $\widetilde{Q}_m = 0$ as we mentioned.\\
	\hfill\\
	Varying $\abs{\widetilde{Q}_m}$ within the range $[0,\frac{\alpha \, l \, \phi_e^2}{\sqrt{3}})$ does not change the characteristics of the phase diagram but affects its shape. Decreasing $\abs{\widetilde{Q}_m}$ shrinks the interval $[T_{c_1},T_{c_2}]$ until $T_{c_1} = T_{c_2}$ at $\widetilde{Q}_m=0$. On the other hand, increasing $\abs{\widetilde{Q}_m}$ widens the interval $[T_{c_1},T_{c_2}]$ with $T_{c_2}\longrightarrow 0$ as $\abs{\widetilde{Q}_m} \longrightarrow \frac{\alpha \, l \, \phi_e^2}{\sqrt{3}}$.
	
	\subsubsection*{Case II.II: $\abs{\widetilde{Q}_m} = \frac{\alpha \, l \, \phi_e^2}{\sqrt{3}}$}
\begin{figure}[htp]
		\centering
		\includegraphics[width=0.8 \textwidth]{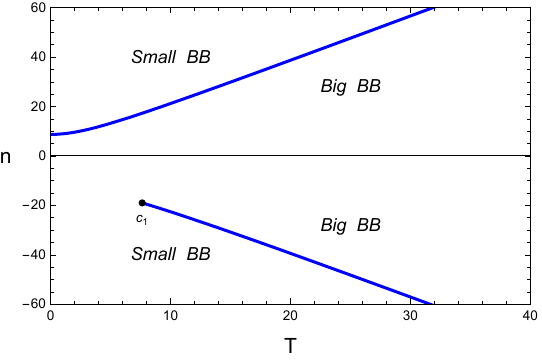}
		\caption{\footnotesize The $n-T$ phase diagram for case II.II. The values used to draw this graph are ($l = 1$, $\phi_e = 15$, $\alpha = \frac{\sqrt{3}}{5}$, $\widetilde{Q}_m = 45$).}
		\label {AII_Q45_N-T}
	\end{figure}

	Substituting the value $\widetilde{Q}_m = \frac{\alpha \, l \, \phi_e^2}{\sqrt{3}}$ in the general formula of $T_h$ in Eq.(\ref{T-Q_m}) we get
	\begin{equation} \label{T-Q_m-caseII.II}
		T_h \,=\frac{3 \left( r_h^{2}+n^2\right)}{4\pi l^2 r_h}-\frac{ 3 r_h^2 \phi_e^2 + \alpha^2 \phi_e^2 \left(  \, l \, \phi_e -  \sqrt{3} n  \right)^2}{12\pi r^3_h}.
	\end{equation}
	\begin{figure}[htp]
		\centering
		\includegraphics[width=0.8 \textwidth]{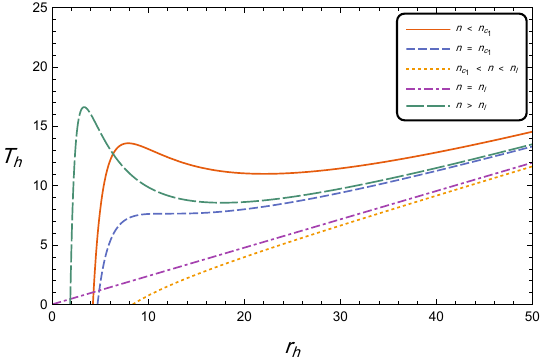}
		\caption{\footnotesize The black brane horizon temperature as a function of the horizon radius in case II.II for different values of $n$. The values used to draw this graph are ($l = 1$, $\phi_e = 15$, $\alpha = \frac{\sqrt{3}}{5}$, $\widetilde{Q}_m = 45$).}
		\label {AII-Q45_T-r}
	\end{figure}
	\hfill\\
    Following our analysis for the critical points we find that we have only one critical point.
	\begin{equation} \label{Nc-case_2_2}
		n_c= - \frac{l \phi_e}{3} \left(6\alpha+\sqrt{3}\right),
	\end{equation}\\
	corresponding to
	\begin{equation} \label{rc-case_2_2II}
		r_{h(c)} =\frac{l \phi_e\sqrt{6 \alpha \left(3 \alpha+\sqrt{3}\right)}}{3},
	\end{equation}\\
	and
	\begin{equation*} \label{Tc-case_2_2I}
		T_{c} =\frac{2 \phi_e \sqrt{6 \alpha \left(3 \alpha+\sqrt{3}\right)}}{3 \pi l},
	\end{equation*}\\
	$n_{c_2}$ is no longer a critical point. However,  it still separates between a monotonic and a non-monotonic $T_h(r_h)$. We will call it $n_l$, where
	\begin{equation} \label{Nl-case_2_2}
		n_l= \frac{l \phi_e}{\sqrt{3}}
	\end{equation}\\
	The reason $n_l$ is not a critical value of $n$ is that $r_{h(c_2)} = 0$ under the given conditions. This is indicated from Fig. \ref{AII-Q45_T-r} where at $n = n_l$, the $T_h-r_h$ curve doesn't have any saddle point. Instead, the horizon temperature of the black brane becomes a linear function of its horizon radius, $T_h=\frac{3 r_h}{4 \pi l^2}$. That results in one of the phases being disjointed from the rest of the phases as seen in the phase diagram in Fig. \ref{AII_Q45_N-T}.
	\hfill\\
	
	\subsubsection*{Case II.III: $\abs{\widetilde{Q}_m} > \frac{\alpha \, l \, \phi_e^2}{\sqrt{3}}$}
\begin{figure*}[htp]
		\centering
		\begin{subfigure}[htp]{0.4\textwidth}   
			\centering 
			\includegraphics[width=\textwidth]{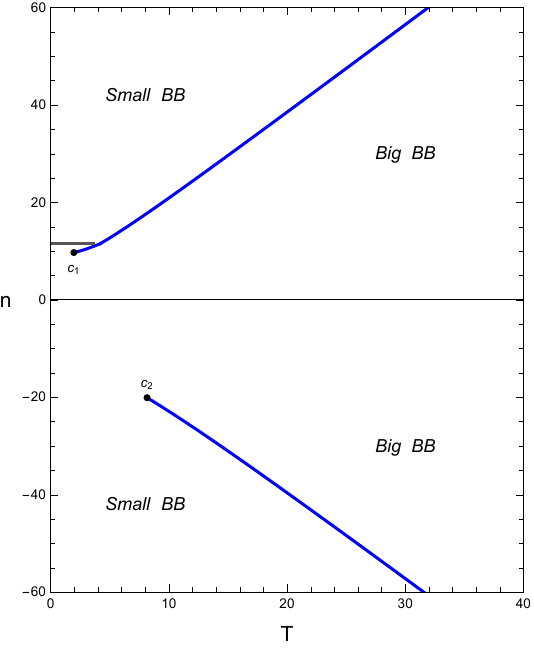}
			\caption[]%
			{{}}    
		\end{subfigure}
		\hfill
		\begin{subfigure}[htp]{0.4\textwidth}   
			\centering 
			\includegraphics[width=\textwidth]{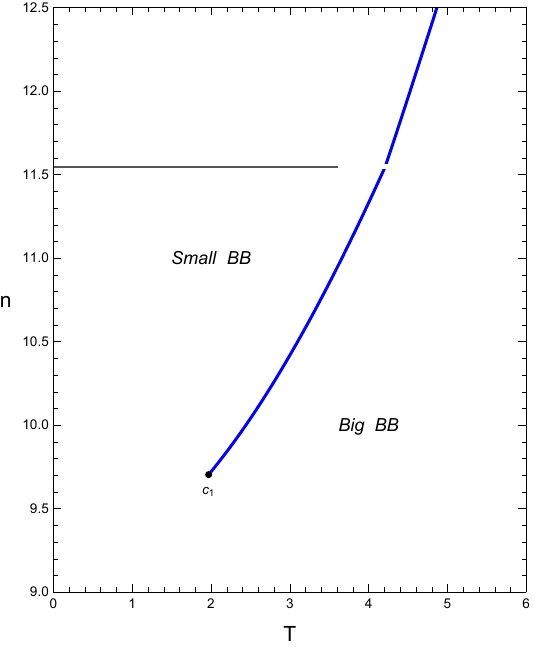}
			\caption[]%
			{{}}    
		\end{subfigure}
		\caption{\footnotesize The $n-T$ phase diagram for case II.III. The values used to draw this graph are ($l = 1$, $\phi_e = 15$, $\alpha = \frac{\sqrt{3}}{5}$, $\widetilde{Q}_m = 60$). The gray line at $n=n_{hp}$ shows the region where black brane cannot exist as a phase.} 
		\label{AII_Q60_N-T}
	\end{figure*}

	This case is very similar to the canonical ensemble of Reissner-Nordstr\"{o}m-anti-de Sitter (RNAdS) black hole \cite{Chamblin:1999hg, Chamblin:1999tk}.\\
	\begin{figure*}[htp]
		\centering
		\begin{subfigure}[htp]{0.49\textwidth}   
			\centering 
			\includegraphics[width=1 \textwidth]{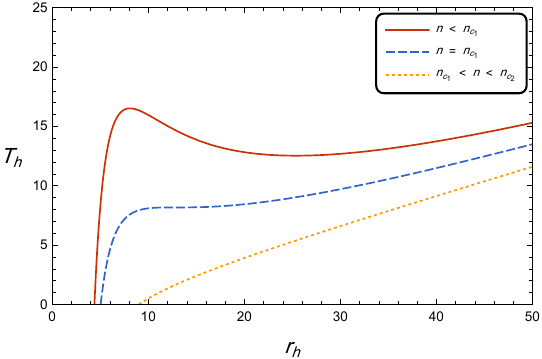}
			\caption[]%
			{{\small $-\infty<n<n_{c_2}$}}    
		\end{subfigure}
		\hfill
		\begin{subfigure}[htp]{0.49\textwidth}   
			\centering 
			\includegraphics[width=1 \textwidth]{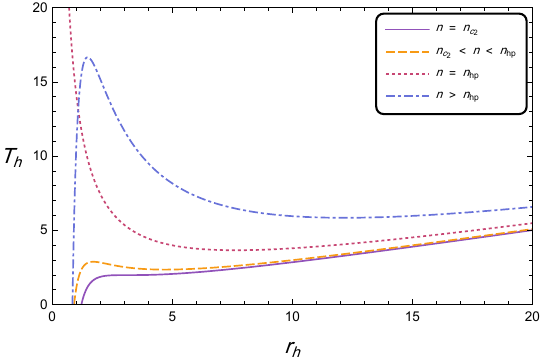}
			\caption[]%
			{{\small $n_{c_2}\leq n<\infty$}}    
		\end{subfigure}
		\caption{\footnotesize The black brane horizon temperature as a function of the horizon radius in case II.III for different values of $n$. The values used to draw this graph are ($l = 1$, $\phi_e = 15$, $\alpha = \frac{\sqrt{3}}{5}$, $\widetilde{Q}_m = 60$). Here we split the graph into two for clarity.}
		\label {AII_Q60_T-r}
	\end{figure*}
	\begin{figure}[htp]
		\centering
	\end{figure}
	\hfill\\
    Referring to our analysis, only two critical points will survive, $n_{c_1}$ and $n_{c_2}$,
	\begin{equation*} \label{Nc-case_2_3I}
		n_{c_1}= -  \alpha l \phi_e - \frac{\sqrt{ 3 l  \left(l \phi_e^2\, (3\alpha^2+1) + 6 \widetilde{Q}_m \right)}}{3},
	\end{equation*}\\
	\begin{equation} \label{Nc-case_2_3II}
		n_{c_2}= -\alpha l \phi_e + \frac{\sqrt{ 3 l  \left(l \phi_e^2\, (3\alpha^2+1) + 6 \widetilde{Q}_m \right)}}{3},
	\end{equation}\\
	corresponding to
	\begin{equation*} \label{rc-case_2_3I}
		r_{h(c_1)} =\frac{\sqrt{9\left(\alpha^2l ^2 \phi_e^2+l\widetilde{Q}_m \right)+3 l \sqrt{3\alpha^2 l \phi_e^2 \left( l \phi_e^2\,(3\alpha^2+1)+6 \widetilde{Q}_m\right)}}}{3},
	\end{equation*}\\
	\begin{equation} \label{rc-case_2_3II}
		r_{h(c_2)} =\frac{\sqrt{9\left(\alpha^2l ^2 \phi_e^2+l\widetilde{Q}_m \right)-3 l \sqrt{3\alpha^2 l \phi_e^2 \left( l \phi_e^2\,(3\alpha^2+1)+6 \widetilde{Q}_m\right)}}}{3},
	\end{equation}\\
	with
	\begin{equation*} \label{Tc-case_2_3I}
		T_{c_1} =\frac{2}{3 \pi l^2} \sqrt{9\left(\alpha^2l ^2 \phi_e^2+l\widetilde{Q}_m \right)+3 l \sqrt{3\alpha^2 l \phi_e^2 \left( l \phi_e^2\,(3\alpha^2+1)+6 \widetilde{Q}_m\right)}},
	\end{equation*}\\
	\begin{equation} \label{Tc-case_2_3II}
		T_{c_2} =\frac{2}{3 \pi l^2}\sqrt{9\left(\alpha^2l ^2 \phi_e^2+l\widetilde{Q}_m \right)-3 l \sqrt{3\alpha^2 l \phi_e^2 \left( l \phi_e^2\,(3\alpha^2+1)+6 \widetilde{Q}_m\right)}},
	\end{equation}\\
	with $T_h(r_h)$ being monotonic for $n \in [n_{c_1}, n_{c_2}]$ and non-monotonic for $n<n_{c_1}$ or $n>n_{c_2}$, with three branches, except for $n=\frac{\widetilde{Q}_m}{\alpha \phi_e}$ where the solution behaves exactly like a Schwarzschild anti-de Sitter solution described in \cite{Hawking:1982dh}. Therefore, We will give this value of $n$ a special name, $n_{hp}$, where hp stands for Hawking-Page.
	At this value of $n$, the horizon temperature has a minimum value $T_{min}$;
       \begin{equation}
      \label{Qhp_T_min}
		T_{min} \,=\frac{\sqrt{9\widetilde{Q}_m^2-3 \alpha^3 l^2 \phi_e^4}}{2\pi \alpha \phi_e l^2}.
	\end{equation}
	with
	\begin{equation} \label{Qhp_r_min}
		r_{min}= \frac{\sqrt{9\widetilde{Q}_m^2-3 \alpha^3 l^2 \phi_e^4}}{3 \alpha \phi_e},
	\end{equation}\\
	For $T>T_{min}$, one thermally stable and one thermally unstable black brane exist. For $T=T_{min}$, only one thermally stable black brane phase exists, and no black branes phase exist for $T<T_{min}$. In fact, as we discussed in case I.I, no known solution with the same asymptotic metric can exist at this range of temperatures. To check the thermal preferability of the phases, we plot $\Omega(T_h)$, Fig \ref{AII_Q60_G-T}.\\
	\begin{figure*}[htp]
		\centering
		\begin{subfigure}[htp]{0.49\textwidth}   
			\centering 
			\includegraphics[width=\textwidth]{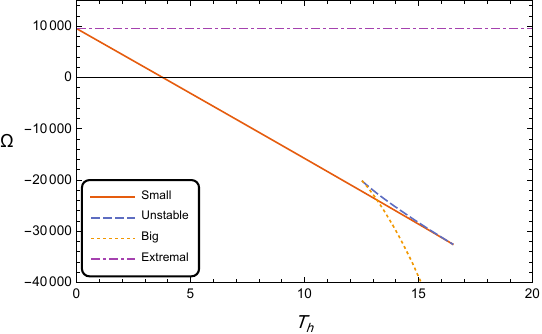}
			\caption[]%
			{{\small $n<n_{c_1}$}}    
		\end{subfigure}
		\hfill
		\begin{subfigure}[htp]{0.49\textwidth}   
			\centering 
			\includegraphics[width=\textwidth]{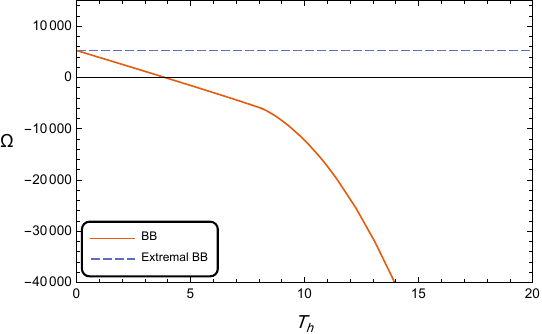}
			\caption[]%
			{{\small $n=n_{c_1}$}}    
		\end{subfigure}
		\vskip\baselineskip
		\begin{subfigure}[htp]{0.49\textwidth}   
			\centering 
			\includegraphics[width=\textwidth]{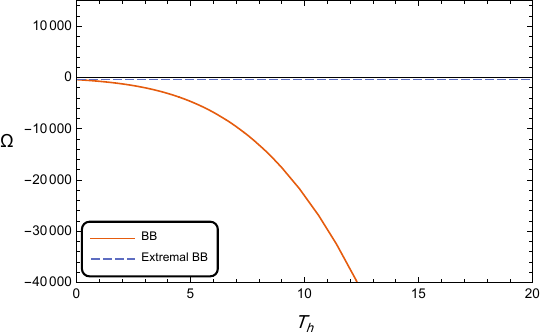}
			\caption[]%
			{{\small $n_{c_1}<n<n_{c_2}$}}    
		\end{subfigure}
		\hfill
		\begin{subfigure}[htp]{0.49\textwidth}   
			\centering 
			\includegraphics[width=\textwidth]{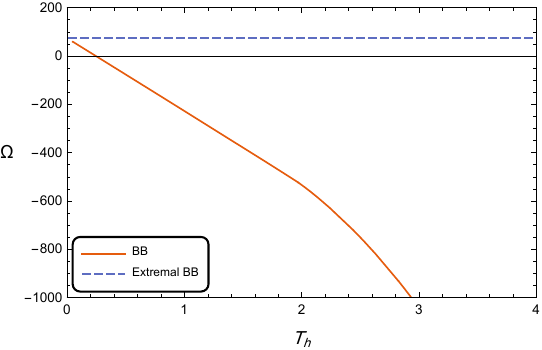}
			\caption[]%
			{{\small $n=n_{c_2}$}}    
		\end{subfigure}
		\vskip\baselineskip
		\begin{subfigure}[htp]{0.49\textwidth}   
			\centering 
			\includegraphics[width=\textwidth]{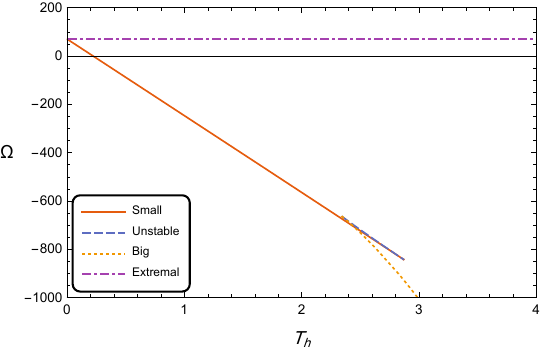}
			\caption[]%
			{{\small $n_{c_2}<n<n_{hp}$}}    
		\end{subfigure}
		\hfill
		\begin{subfigure}[htp]{0.49\textwidth}   
			\centering 
			\includegraphics[width=\textwidth]{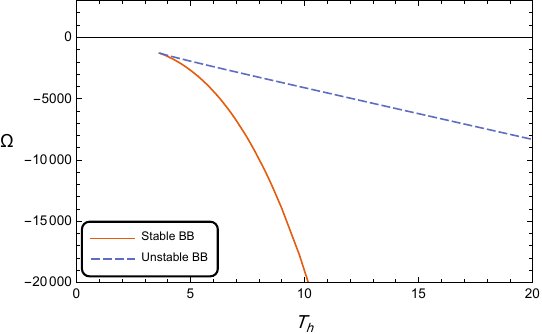}
			\caption[]%
			{{\small $n = n_{hp}$}}    
		\end{subfigure}
	\end{figure*}
	\begin{figure*}
		\ContinuedFloat
		\centering 
		\begin{subfigure}[htp]{0.49\textwidth}   
			\centering 
			\includegraphics[width=\textwidth]{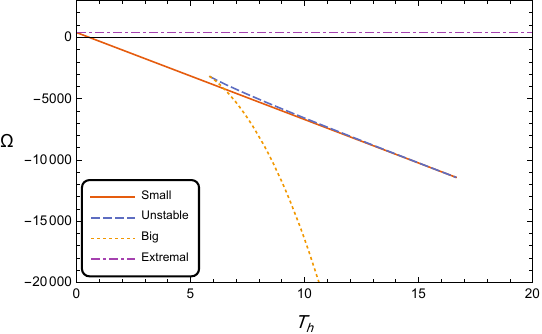}
			\caption[]%
			{{\small $n > n_{hp}$}}    
		\end{subfigure}
		\caption{\footnotesize The grand potential density of the black brane as a function of the horizon temperature in case II.III for different values of $n$. The values used to draw this graph are ($l = 1$, $\phi_e = 15$, $\alpha = \frac{\sqrt{3}}{5}$, $\widetilde{Q}_m = 60$).}
		\label{AII_Q60_G-T}
	\end{figure*}
	\newpage
	Summarizing our results, we have:
	\begin{itemize}
		\item For $\,n < n_{c_1}$,  $\;n \in (n_{c_2},n_{hp})$, or  $\;n > n_{hp}\,$  and $\,T<T_{min}$ or $T>T_{max}\,$, the only stable phase is one black brane.
		\item For $\, n < n_{c_1}$,  $\; n \in (n_{c_2}, n_{hp})$, or  $\; n > n_{hp}\,$ and $T \in [T_{min}, T_{tr})$, there are two stable black brane phases. The small black brane is thermodynamically preferred, while the big black brane is metastable.
		\item For $\, n < n_{c_1}$,  $\; n \in (n_{c_2}, n_{hp})$, or  $\; n > n_{hp}\,$ and $T= T_{tr}$, first-order phase transition occurs between the small and the big black branes.
		\item For $\, n < n_{c_1}$,  $\; n \in (n_{c_2}, n_{hp})$, or  $\; n > n_{hp}\,$ and $T \in (T_{tr}, T_{max}]$, there are two stable black brane phases. The big black brane is thermodynamically preferred while the small black brane is metastable.
		\item For $n = n_{c_1}$ and $T = T_{c_1}$ or $n = n_{c_2}$ and $T = T_{c_2}$, there is a critical point where the first-order phase transition between the small and the big black branes ends, and the two branes become indistinguishable.
		\item For $n \in [n_{c_1},n_{c_2}]$, the only stable phase is one black brane.
		\item  For $n = n_{hp}$ and $T < T_{min}$, there are no possible black brane phases.
		\item  For $n = n_{hp}$ and $T \geq T_{min}$, the only stable phase is a one black brane.
	\end{itemize}
	The complete phase diagram is shown in Fig \ref{AII_Q60_N-T}.\\
	\break
	\\
	\hfill\\
	\break
	\subsubsection{Case III: $\abs{\alpha} > \frac{\sqrt{3}}{3}$}
	Similar to Case II, there are no vanishing thermodynamic quantities, so the temperature of the black brane takes the general form in Eq.(\ref{T-Q_m}). Results now depend on the value of $\widetilde{Q}_m$.
	\subsubsection*{Case III.I: $\abs{\widetilde{Q}_m} \in [0,\frac{\alpha \, l \, \phi_e^2}{\sqrt{3}}]$}
	This case is exactly the same as Case II.I
	\subsubsection*{Case III.II: $\abs{\widetilde{Q}_m} \in (\frac{\alpha \, l \, \phi_e^2}{\sqrt{3}},\frac{l \phi_e^2 \left(1+3\alpha^2\right)}{6}]$ 
	}

\begin{figure*}[htp]
		\centering
		\begin{subfigure}[htp]{0.4\textwidth}   
			\centering 
			\includegraphics[width=\textwidth]{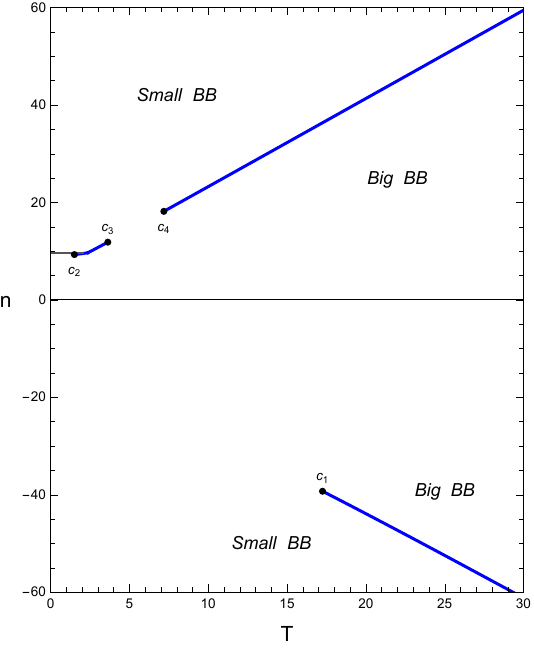}
			\caption[]%
			{{}}    
		\end{subfigure}
		\hfill
		\begin{subfigure}[htp]{0.4\textwidth}   
			\centering 
			\includegraphics[width=\textwidth]{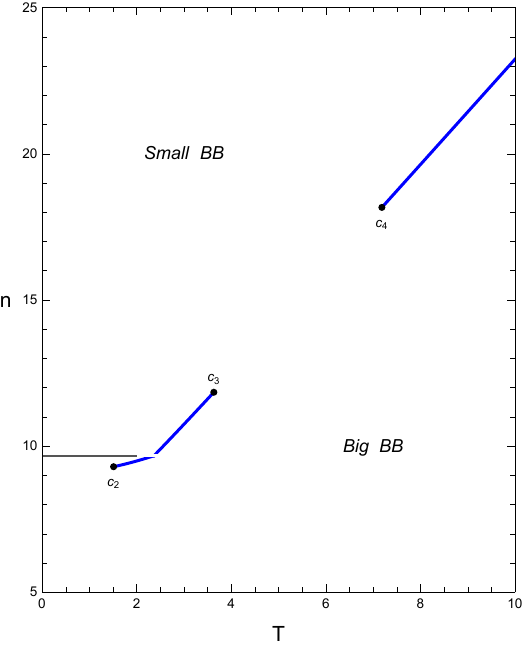}
			\caption[]%
			{{}}    
		\end{subfigure}
		\caption{\footnotesize The $n-T$ phase diagram for case III.II. The values used to draw this graph are ($l = 1$, $\phi_e = 15$, $\alpha = 1$, $\widetilde{Q}_m = 145$). The gray line at $n=n_{hp}$ shows the region where black brane cannot exist as a phase.} 
		\label{AIII-Q145_N-T}
	\end{figure*}

	As stated earlier, the monotonicity of $T_h(r_h)$  varies due to the value of $n$. In this case, we have all the four critical points 
    \begin{equation} \label{Nc-gIc}
		n_{c_{1,2}}= -  \alpha l \phi_e \mp \frac{1}{3}{\sqrt{ 3 l  \left(l \phi_e^2\, (3\alpha^2+1) + 6 \widetilde{Q}_m \right)}},
	\end{equation}\\
    corresponding to
    \begin{equation} \label{rc12-gIc}
		r_{h(c_{1, 2})} =\frac{1}{3}{\sqrt{9\left(\alpha^2l ^2 \phi_e^2+l\widetilde{Q}_m \right) \pm 3 l \sqrt{3\alpha^2 l \phi_e^2 \left( l \phi_e^2\,(3\alpha^2+1)+6 \widetilde{Q}_m\right)}}}\,
	\end{equation}\\
    and
    \begin{equation} \label{Tc-case_3_2Ic}
		T_{c_{1,2}} =\frac{2}{3 \pi l^2} \sqrt{9\left(\alpha^2l ^2 \phi_e^2+l\widetilde{Q}_m \right)\pm3 l \sqrt{3\alpha^2 l \phi_e^2 \left( l \phi_e^2\,(3\alpha^2+1)+6 \widetilde{Q}_m\right)}},
	\end{equation}\\
    and
    \begin{equation} \label{Nc-gIIc}
		n_{c_{3,4}}=   \alpha l \phi_e \mp \frac{1}{3}{\sqrt{ 3 l  \left(l \phi_e^2\, (3\alpha^2+1) - 6 \widetilde{Q}_m \right)}},
	\end{equation}\\
    corresponding to
    \begin{equation} \label{rc34-gIc}
		r_{h(c_{3, 4})} =\frac{1}{3}{\sqrt{9\left(\alpha^2l ^2 \phi_e^2-l\widetilde{Q}_m \right) \mp 3 l \sqrt{3\alpha^2 l \phi_e^2 \left( l \phi_e^2\,(3\alpha^2+1)-6 \widetilde{Q}_m\right)}}},
	\end{equation}\\
    and
	\begin{equation} \label{Tc-case_3_2IIIc}
		T_{c_{3,4}} =\frac{2}{3 \pi l^2}\sqrt{9\left(\alpha^2l ^2 \phi_e^2-l\widetilde{Q}_m \right)\mp3 l \sqrt{3\alpha^2 l \phi_e^2 \left( l \phi_e^2\,(3\alpha^2+1)-6 \widetilde{Q}_m\right)}},
	\end{equation}\\
	\begin{figure*}[htp]
		\centering
		\begin{subfigure}[htp]{0.49\textwidth}   
			\centering 
			\includegraphics[width=1 \textwidth]{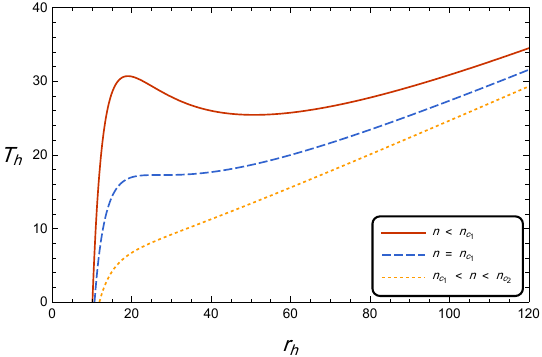}
			\caption[]%
			{{\small $-\infty < n < n_{c_2}$}}    
		\end{subfigure}
		\hfill
		\begin{subfigure}[htp]{0.49\textwidth}   
			\centering 
			\includegraphics[width=1 \textwidth]{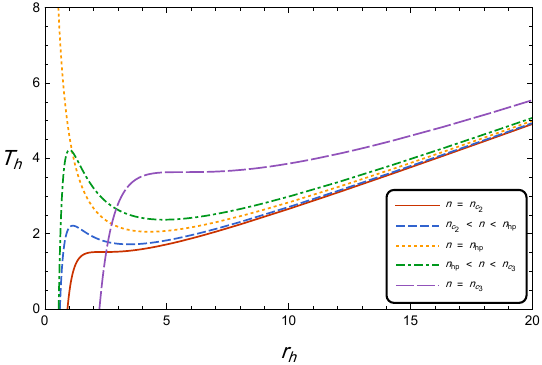}
			\caption[]%
			{{\small $n_{c_2} \leq n \leq n_{c_3}$}}    
		\end{subfigure}
		\vskip\baselineskip
		\begin{subfigure}[htp]{0.49\textwidth} 
			\centering 
			\includegraphics[width=1 \textwidth]{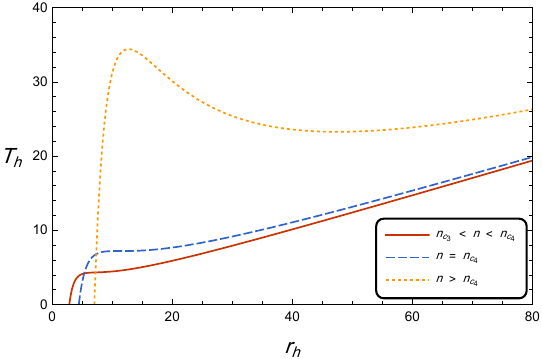}
			\caption[]%
			{{\small $n_{c_3} < n < \infty$}}    
		\end{subfigure}
		\caption{\footnotesize The black brane horizon temperature as a function of the horizon radius in case III.II for different values of $n$. The values used to draw this graph are ($l = 1$, $\phi_e = 15$, $\alpha = 1$, $\widetilde{Q}_m = 145$). Here we split the graph into three for clarity.}
		\label{AIII_Q145_T-r}
	\end{figure*}
	\hfill\\	
	From Fig. \ref{AIII_Q145_T-r} we can see that for $n \in [n_{c_1},n_{c_2}]$ or $n \in [n_{c_3},n_{c_4}]$ the horizon temperature of the black brane, $T_h(r_h)$, is monotonic. For $n < n_{c_1}$ or $n > n_{c_4}$, $T_h(r_h)$ is non-monotonic, having three branches. Finally, for $n \in (n_{c_2},n_{c_3})$, it also has three branches except for $n = n_{hp} = \frac{\widetilde{Q}_m}{\alpha \phi_e}$ where $T_h(r_h)$ has two branches and a minimum value $T_{min}$ given by Eq.(\ref{Qhp_T_min}) at $r_h = r_{min}$ given by Eq.(\ref{Qhp_r_min}).\\
	\hfill\\
	Following the same procedure as in the previous cases, we next check the thermal preferability of the phases by plotting $\Omega(T_h)$, Fig. \ref{AIII_Q145_G-T}.\\
	\begin{figure*}[h!]
		\centering
		\begin{subfigure}[htp]{0.49\textwidth}   
			\centering 
			\includegraphics[width=\textwidth]{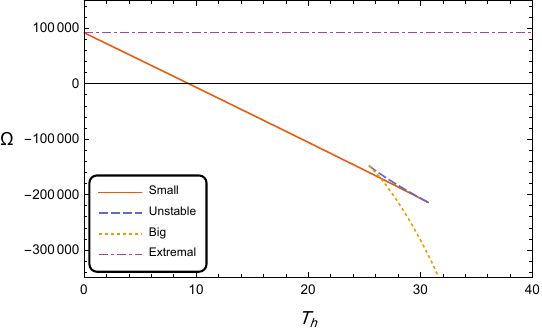}
			\caption[]%
			{{\small $n<n_{c_1}$}}    
		\end{subfigure}
		\hfill
		\begin{subfigure}[htp]{0.49\textwidth}   
			\centering 
			\includegraphics[width=\textwidth]{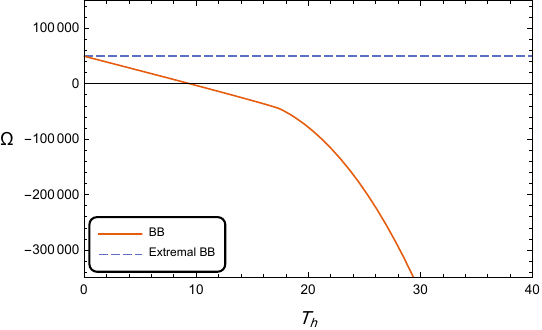}
			\caption[]%
			{{\small $n=n_{c_1}$}}    
		\end{subfigure}
		\vskip\baselineskip
		\begin{subfigure}[htp]{0.49\textwidth}   
			\centering 
			\includegraphics[width=\textwidth]{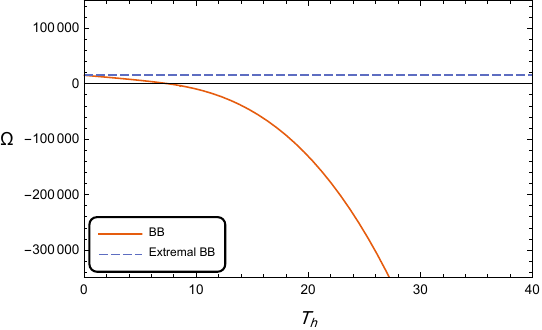}
			\caption[]%
			{{\small $n_{c_1}<n<n_{c_2}$}}    
		\end{subfigure}
		\hfill
		\begin{subfigure}[htp]{0.49\textwidth}   
			\centering 
			\includegraphics[width=\textwidth]{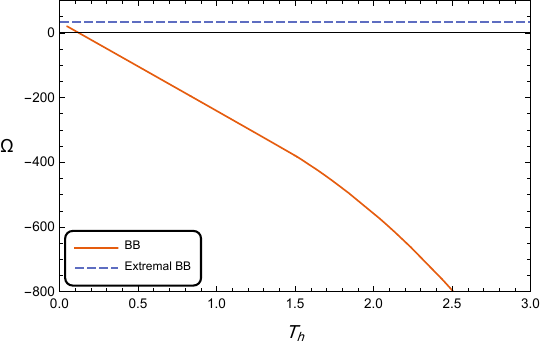}
			\caption[]%
			{{\small $n=n_{c_2}$}}    
		\end{subfigure}
		\vskip\baselineskip
		\begin{subfigure}[htp]{0.49\textwidth}   
			\centering 
			\includegraphics[width=\textwidth]{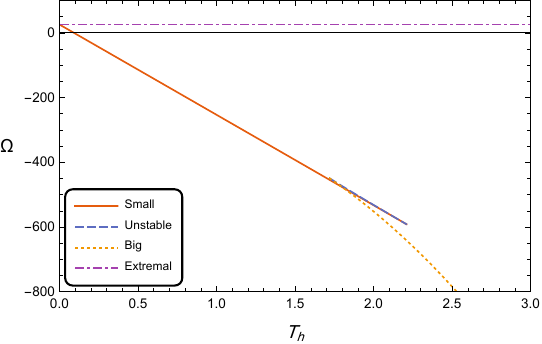}
			\caption[]%
			{{\small $n_{c_2}<n<n_{hp}$}}    
		\end{subfigure}
		\hfill
		\begin{subfigure}[htp]{0.49\textwidth}   
			\centering 
			\includegraphics[width=\textwidth]{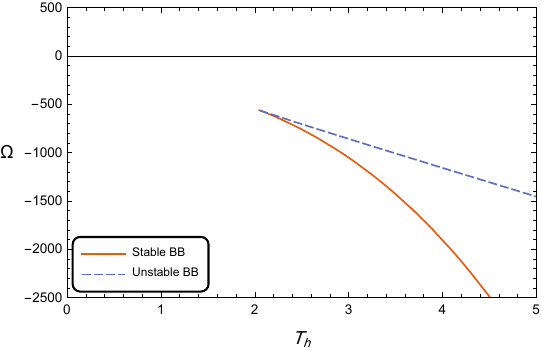}
			\caption[]%
			{{\small $n = n_{hp}$}}    
		\end{subfigure}
	\end{figure*}
	\begin{figure*}[h!]
		\ContinuedFloat
		\centering
		\begin{subfigure}[htp]{0.49\textwidth}   
			\centering 
			\includegraphics[width=\textwidth]{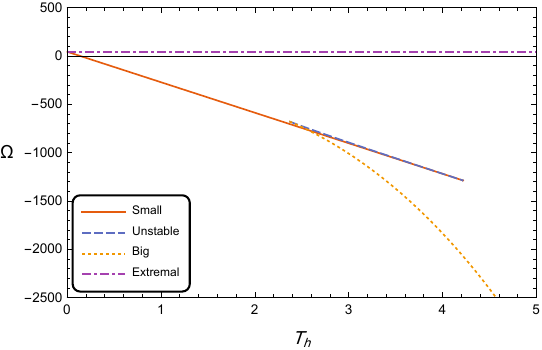}
			\caption[]%
			{{\small $n_{hp}<n<n_{c_3}$}}    
		\end{subfigure}
		\hfill
		\begin{subfigure}[htp]{0.49\textwidth}   
			\centering 
			\includegraphics[width=\textwidth]{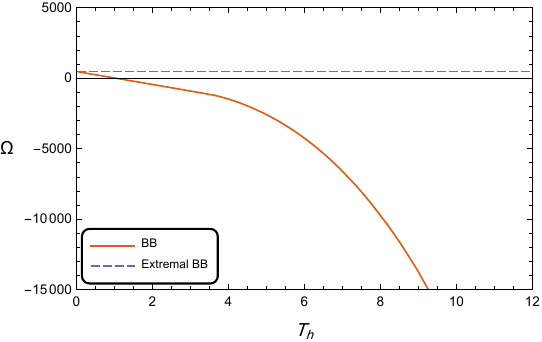}
			\caption[]%
			{{\small $n=n_{c_3}$}}    
		\end{subfigure}
		\vskip\baselineskip
		\begin{subfigure}[htp]{0.49\textwidth}   
			\centering 
			\includegraphics[width=\textwidth]{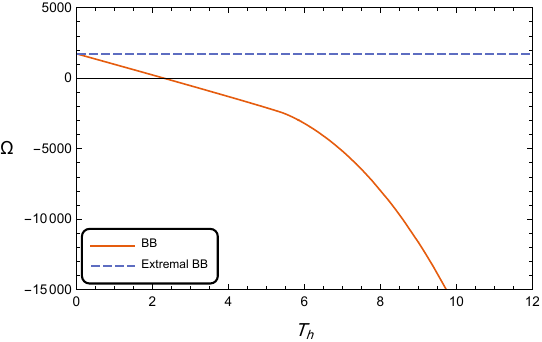}
			\caption[]%
			{{\small $n_{c_3}<n<n_{c_4}$}}    
		\end{subfigure}
		\hfill
		\begin{subfigure}[htp]{0.49\textwidth}   
			\centering 
			\includegraphics[width=\textwidth]{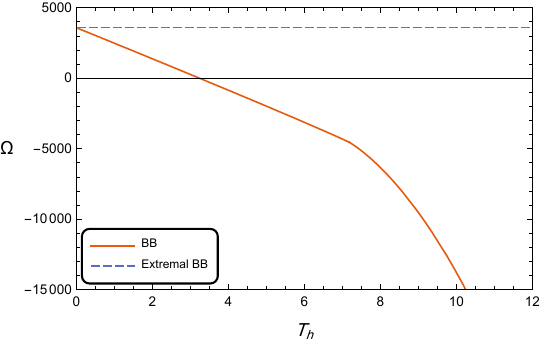}
			\caption[]%
			{{\small $n=n_{c_4}$}}    
		\end{subfigure}
		\vskip\baselineskip
		\begin{subfigure}[htp]{0.49\textwidth}   
			\centering 
			\includegraphics[width=\textwidth]{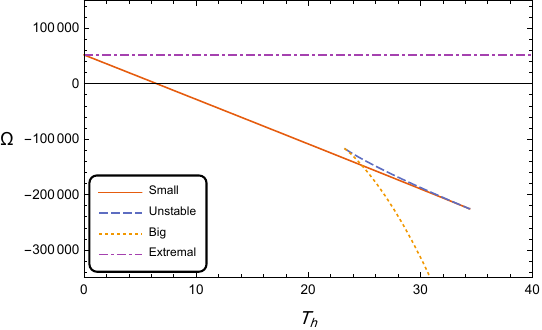}
			\caption[]%
			{{\small $n > n_{c_4}$}}    
		\end{subfigure}
		\caption{\footnotesize The grand potential density of the black brane as a function of its horizon temperature in case III.II for different values of $n$. The values used to draw this graph are ($l = 1$, $\phi_e = 15$, $\alpha = 1$, $\widetilde{Q}_m = 145$).}
		\label{AIII_Q145_G-T}
	\end{figure*}
	\hfil\\
    \newpage
	In summary, we have
	\begin{itemize}
		\item For $\,n < n_{c_1}$,  $\;n \in (n_{c_2},n_{hp})$, $\;n \in (n_{hp},n_{c_3})$, or  $\;n > n_{c_4}\,$  and $\,T<T_{min}$ or $T>T_{max}\,$, the only stable phase is one black brane.
		\item For  $\, n < n_{c_1}$,  $\; n \in (n_{c_2}, n_{hp})$, $\; n \in (n_{hp}, n_{c_3})$, or  $\; n > n_{c_4}\,$  and $T \in [T_{min}, T_{tr})$, there are two stable black brane phases, the small black brane is thermodynamically preferred while the big black brane is metastable.
		\item For $\,n < n_{c_1}$,  $\;n \in (n_{c_2}, n_{hp})$, $\;n \in (n_{hp}, n_{c_3})$, or  $\;n > n_{c_4}\,$  and $T= T_{tr}$, first-order phase transition occurs between the small and the big black branes.
		\item For $\,n < n_{c_1}$,  $\;n \in (n_{c_2}, n_{hp})$, $\;n \in (n_{hp}, n_{c_3})$, or  $\;n > n_{c_4}\,$   and $T \in (T_{tr}, T_{max}]$, there are two stable black brane phases, the big black brane is thermodynamically preferred while the small black brane is metastable.
		\item For $n = n_{c_1}$  and $T = T_{c_1}$, $n = n_{c_2}$ and $T = T_{c_2}$, $n = n_{c_3}$   and $T = T_{c_3}$ or $n = n_{c_4}$  and $T = T_{c_4}$, there is a critical point where the first-order phase transition between the small and the big black branes ends, and the two branes become indistinguishable.
		\item For $n \in [n_{c_1},n_{c_2}]$ or $n \in [n_{c_3},n_{c_4}]$, the only stable phase is one black brane.
		\item  For $n = n_{hp}$ and $T < T_{min}$, there are no possible black brane phases.
		\item  For $n = n_{hp}$ and $T \geq T_{min}$, the only stable phases is one black brane.
	\end{itemize}
	The complete phase diagram for this case is shown in Fig. \ref{AIII-Q145_N-T}
	
\begin{figure*}[htp]
		\centering
		\begin{subfigure}[htp]{0.4\textwidth}   
			\centering 
			\includegraphics[width=\textwidth]{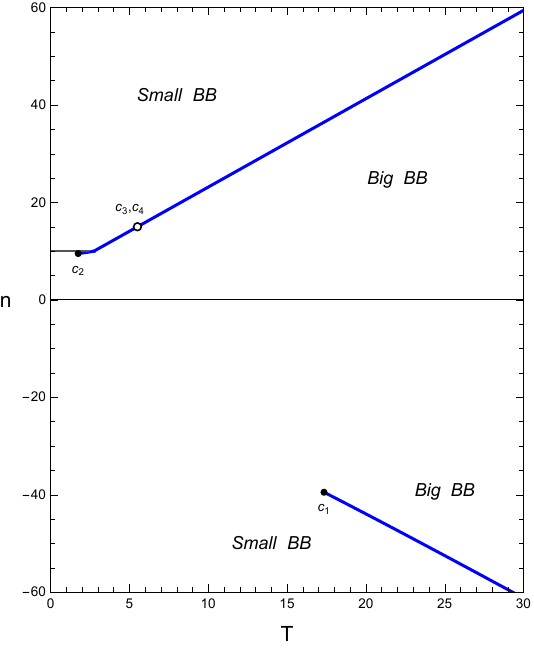}
			\caption[]%
			{{}}    
		\end{subfigure}
		\hfill
		\begin{subfigure}[htp]{0.4\textwidth}   
			\centering 
			\includegraphics[width=\textwidth]{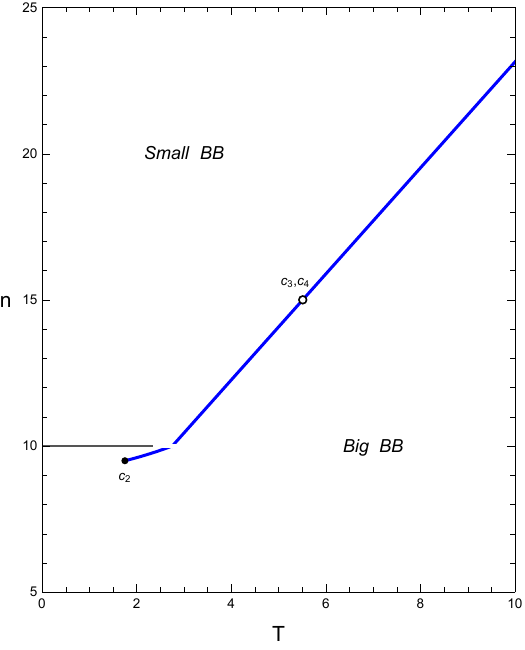}
			\caption[]%
			{{}}    
		\end{subfigure}
		\caption{\footnotesize The $n-T$ phase diagram for case III.II where two of the critical points merge. Values used to draw this graph are ($l = 1$, $\phi_e = 15$, $\alpha = 1$, $\widetilde{Q}_m = 150$) The gray line at $n=n_{hp}$ shows the region where black brane cannot exist as a phase.
        } 
		\label{AIII-Q150_N-T}
	\end{figure*}

	\hfill\\
	As the value of $\widetilde{Q}_m$ increases in the region $ (\frac{\alpha \, l \, \phi_e^2}{\sqrt{3}},\frac{l \phi_e^2 \left(1+3\alpha^2\right)}{6}]$, the intervals $(n_{c_2}, n_{c_3})$ and $(T_{c_2}, T_{c_3})$ shrink until the two critical points merge at $\widetilde{Q}_m = \frac{l \phi_e^2 \left(1+3\alpha^2\right)}{6}$. This can be seen from the phase diagram in Fig. \ref{AIII-Q150_N-T}.\\
	
	\subsubsection*{Case III.III: $\abs{\widetilde{Q}_m} > \frac{l \phi_e^2 \left(1+3\alpha^2\right)}{6}$}
	As the value of $\abs{\widetilde{Q}_m}$ increases to be greater than $\frac{l \phi_e^2 \left(1+3\alpha^2\right)}{6}$ the two merged critical points vanish completely and the case becomes exactly the same as Case II.III.\\
	\pagebreak
	\section{Conclusion}\label{sec4}
	
    We started with a review of the phase structure of the NUT-less dyonic black branes in the mixed ensemble, i.e. an ensemble with fixed electric potential and magnetic charge. Also, we have calculated the thermodynamic quantities of the solution and show that they obey the first law and Gibbs-Duhem relation. We showed that in the NUT-less case the phase structure is trivial since there are no phase transitions and the temperature is a monotonic function of the black brane horizon radius.\\
    \hfill\\
    In the main work of the paper we have applied the thermodynamic treatment introduced in \cite{Awad:2022jgn,Awad:2023lyt} to the dyonic-NUT-AdS spacetimes with flat horizons to obtain various thermodynamic quantities. We showed that these quantities satisfy the first law as well as the Gibbs-Duhem relation. It was intriguing to notice that the first law is satisfied by a larger class of charges and potentials which depends on some additional arbitrary parameter $\alpha$ which in turn affects the phase structure.
    We analyzed the phase structure of these solutions through a mixed ensemble where we described the phase diagrams as NUT parameter-temperature graphs to show possible transitions between big and small black brane phases. We showed that due to the absence of Misner strings, one doesn't need to fix $\alpha$, therefore, it takes any value \cite{Tharwat:2023vku}. Depending on the value of $\alpha$, one can classify the phase study into different distinct cases which could be further split into few sub-cases depending on the value of the other thermodynamic quantities. In most of these cases we have first-order phase transitions which end at one or more critical points. We found that these classes could have up to four critical points, again, depending on $\alpha$. These cases are summarized in Table \ref{table.1}.\\
    \begin{table}[H]
    \centering
    \begin{tabular}{|c|c|c|}
    \hline
    \multirow{2}{*}{Range of $\abs{\alpha}$} & \multirow{2}{*}{Range of $\abs{\widetilde{Q}_m}$} & \multirow{2}{*}{Number of Critical Points} \\
     &  &  \\ \hline
    \multirow{2}{*}{$= 0$} & $= 0$ & 0 \\ \cline{2-3} 
     & $\neq 0$ & 2 \\ \hline
    \multirow{3}{*}{$\in (0,\frac{\sqrt{3}}{3}]$} & $< \frac{\alpha \, l \, \phi_e^2}{\sqrt{3}}$ & 2 \\ \cline{2-3} 
     & $= \frac{\alpha \, l \, \phi_e^2}{\sqrt{3}}$ & 1 \\ \cline{2-3} 
     & $> \frac{\alpha \, l \, \phi_e^2}{\sqrt{3}}$ & 2 \\ \hline
    \multirow{3}{*}{$> \frac{\sqrt{3}}{3}$} & $\in [0,\frac{\alpha \, l \, \phi_e^2}{\sqrt{3}}]$ & 2 \\ \cline{2-3} 
     & $\in (\frac{\alpha \, l \, \phi_e^2}{\sqrt{3}},\frac{l \phi_e^2 \left(1+3\alpha^2\right)}{6}]$ & 4 \\ \cline{2-3} 
     & $> \frac{l \phi_e^2 \left(1+3\alpha^2\right)}{6}$ & 2 \\ \hline
    \end{tabular}
    \caption{Number of thermodynamic critical points for the dyonic Taub–NUT–AdS black brane under different parameter ranges of $\alpha$ and $\widetilde{Q}_m$, corresponding to the sub-cases analyzed in Section 3.}
    \label{table.1}
    \end{table}
    \hfill\\
    It is important to notice that these phase structures are qualitatively different from the van der Waals structure of the liquid-gas phase transition.\\
    \hfill\\    
    One possible extension to this work is to study the Kerr-Newman-NUT in Minkowski as well as Anti-de Sitter spaces, which could enrich our understanding of the thermodynamics of the NUT-AdS spaces and their phase structure. Another extension of this work could be investigating the Plebanski-Demianski metric, a class of solutions that depends on seven parameters: mass, rotation, electric and magnetic charges, nut, acceleration, and cosmological constant. We hope that we can study one of these spacetimes in another work in the near future. 

	\bibliographystyle{apsrev}             
	\bibliography{ref}
\end{document}